\documentclass[11pt]{article}
\pdfoutput=1

\usepackage{amsmath}
\usepackage{amssymb}
\usepackage{mathrsfs}
\usepackage{graphicx}
\usepackage{enumerate}
\usepackage[font=small]{caption}
\usepackage{subcaption}
\usepackage[percent]{overpic}
\usepackage[utf8]{inputenc}
\usepackage{amsfonts}
\usepackage{mathtools}
\usepackage{braket}
\usepackage{bbm}
\usepackage{slashed}
\usepackage{enumitem}
\usepackage[usenames, dvipsnames]{color}
\usepackage[UKenglish]{isodate}
\usepackage[dvipsnames]{xcolor}
\usepackage[normalem]{ulem}
\usepackage{csquotes}
\usepackage{outlines}
\usepackage{jheppub}

\newcommand\bb{\begin{equation*}}
\newcommand\be{\begin{equation*}}
\newcommand\bne{\begin{equation}}
\newcommand\bse{\begin{equation*} \begin{split}}
\newcommand\ee{\end{equation*}}
\newcommand\ene{\end{equation}}
\newcommand\ese{\end{equation*} \end{split}}
\newcommand\bfr{\begin{flushright}}
\newcommand\efr{\end{flushright}}
\newcommand\bc{\begin{center}}
\newcommand\ec{\end{center}}

\newcommand\bi{\begin{itemize}[leftmargin=*]}
\newcommand\ei{\end{itemize}}
\newcommand{\sss}[1]{\subsubsection*{#1}}

\newcommand{\op}{\mathcal{O}}

\newcommand{\del}{\partial}

\newcommand{\fn}[1]{\footnote{#1}}

\newcommand{\TCA}{\tilde{\mathcal{A}}}

\newcommand{\pd}[2]{\frac{\partial #1}{\partial #2}}
\DeclareMathOperator{\area}{area}

\makeatletter
\def\@fpheader{\relax}
\makeatother

\numberwithin{equation}{section}

\title{Bit threads in higher-curvature gravity}

\author[a]{Jonathan Harper,}
\author[a,b]{Matthew Headrick,}
\author[a]{and Andrew Rolph}
\affiliation[a]{Martin Fisher School of Physics, Brandeis University, Waltham MA 02453, USA}
\affiliation[b]{Center for Theoretical Physics, Massachusetts Institute of Technology, Cambridge, MA 02139, USA }
\preprint{BRX-TH-6327, MIT-CTP/5001}

\abstract{We generalize holographic bit threads to bulk theories with a gravitational action containing higher-curvature terms. Bit threads are a reformulation of holographic entanglement entropy, where the entropy is given by the maximum number of threads emanating from a boundary region into the bulk. We show that the addition of higher-curvature terms adds corrections to the bit thread thickness that depend on the local geometry and thread orientation. Two different methods are given: determination of the density bound by requiring the maximum number of threads through a given surface to reproduce the entanglement entropy functional on that surface, and application of Lagrange dualization. The results of the two methods are applied to Gauss-Bonnet gravity as the simplest non-trivial example.
}

\begin{document}

\maketitle

\section{Introduction}
Holographic entanglement entropy (HEE) has had a profound impact on our understanding of quantum gravity, by directly connecting quantum information and geometry. According to the Ryu-Takayanagi (RT) formula \cite{Ryu:2006bv}, the entropy of a boundary region $A$ is given by the area of the minimal-area bulk surface homologous to $A$:
\begin{equation} \label{eq:RT formula}
S(A) = \frac1{4G_{\rm N}}\min_{m\sim A}\area(m).
\end{equation}
Bit threads \cite{Freedman:2016zud} are a reformulation of the RT formula introduced to address some of the conceptual issues arising from the RT formula. To understand them, one first defines a \emph{flow}, a vector field $v$ satisfying $\nabla_\mu v^\mu=0$, $|v|\le1$ everywhere. A consequence of the divergencelessness is that the flux of $v$ through $A$ equals the flux through any surface homologous to $A$. Maximizing the flux picks out the minimal surface as the ``bottleneck'', on which the norm bound is saturated, giving a flux equal to the minimal surface area. A simple analogy is determining the size of a water pipe's bottleneck by maximizing the flow through one end. Mathematically, this is expressed in the max flow-min cut theorem,
\bne\label{eq:mfmc}
\min_{m\sim A} \text{area}(m) = \max_v \int_A \sqrt h\,n_\mu v^\mu \quad\text{where}\quad \nabla_\mu v^\mu = 0\,,\quad |v| \leq 1\,.
\ene
Based on this theorem, we can rewrite the RT formula as
\begin{equation}\label{eq:flow}
S(A) = \frac1{4G_{\rm N}}\max_v\int_A \sqrt h\,n_\mu v^\mu\,.
\end{equation}

A bit thread is an integral curve of $v$. The threads are chosen as a subset of the integral curves with transverse density equal to $|v|/4G_{\rm N}$; effectively, each thread has cross-sectional area $4G_N$. The fact that threads have thickness and cannot intersect then gives an interpretation to the vector field norm bound: there is a limit to how closely the threads can be packed together. For configurations that maximize the number of threads out of $A$, the threads are maximally packed on and directionally normal to the minimal surface, which acts as a bottleneck. There is a redundancy in which discrete members in the continuous family of integral curves are given a cross-sectional area and called bit threads. At AdS scales the bit thread number density is of the order $N^2$, so the discrete family of bit threads is indistinguishable from the continuous family of integral curves, and the terms ``bit thread'' and ``flow'' can be used interchangeably. The maximum number of threads that can be placed on $A$ gives the entanglement entropy, as if each bit thread connected an EPR pair of qubits between $A$ and its complement region $A^c$; bit threads are thus a layer of quantum information theoretic interpretation on top of the vector field $v$. Unlike the minimal surface, the bit thread configuration changes continuously under continuous deformations of the region $A$. Furthermore, they allow for very natural expressions for important information-theoretic quantities such as the conditional entropy, mutual information, and conditional mutual information. They also provide proofs of important properties like subadditivity and strong subadditivity that correspond directly to the information-theoretic meaning of these properties.

The RT formula \eqref{eq:RT formula}, and therefore the bit-thread formula \eqref{eq:flow}, requires the bulk theory and state to obey a particular set of conditions, including being in the classical limit (large-$N$ limit of the field theory), governed by Einstein gravity (strong-coupling limit of the field theory), and in a state possessing a time-reflection symmetry. It is by now more or less understood how to relax each of these conditions on the RT formula. If the bit threads indeed have a fundamental physical significance---as opposed to being just a mathematical artifact of the simplicity of the RT formula---then it should be possible to relax these conditions for the bit threads as well.

In this paper, we will take up this challenge by exploring how to relax one of the above conditions, namely the assumption that the bulk is described by Einstein gravity. In other words, we will show how to formulate bit threads when the bulk gravitational action includes higher-curvature terms. This is equivalent to moving away from the strong-coupling limit in the boundary theory. For work on black-hole entropy and HEE in higher-curvature gravity theories see \cite{Bhattacharyya:2013gra,Bhattacharyya:2013jma,Bhattacharyya:2014yga,Jacobson1993,Caceres2017,Erdmenger2014,Chen2013,MohammadiMozaffar2016,Dong:2013qoa,Camps2014,Haehl2017}. The original bit thread paper \cite{Freedman:2016zud} discussed one possible way to generalize bit threads to higher-curvature gravity, by employing a conjectural generalization of the Weyl law for the spectrum of the scalar Laplacian to vector fields. In this paper, we will take a more direct approach.

In higher-curvature theories, the right-hand side of the RT formula is corrected by the minimum of a local geometrical functional of the surface:
\begin{equation} \label{eq:RT formula2}
S(A) = \frac1{4G_{\rm N}}\min_{m\sim A}a_\lambda(m)\,,\qquad
a_\lambda(m):=\int_m\sqrt g\left(1+\lambda\tilde{\mathcal{A}}\right),
\end{equation}
where $\lambda$ is a small parameter and $\tilde A$ is a function of the intrinsic and extrinsic geometry and the ambient metric \cite{Dong:2013qoa}. For example, the simplest example of a higher curvature correction is for Gauss-Bonnet gravity, where $\lambda$ is the coefficient of the Gauss-Bonnet term in the gravitational action and $\TCA$ is the scalar curvature of the induced metric on $m$ \cite{deBoer:2011wk,Hung:2011xb}.

In order to generalize the bit threads to higher curvature gravity, we need to find a generalization of the max flow-min cut theorem \eqref{eq:mfmc} in which the area functional is replaced by $a_\lambda$ and the right-hand side is corrected in an appropriate way. 
The right-hand side 
involves three ingredients: the objective to be maximized $\int_A\sqrt hn_\mu v^\mu$, the divergencelessness condition $\nabla_\mu v^\mu=0$, and the norm bound $|v|\le1$. We will find that the $\lambda \tilde{\mathcal{A}}$ correction can be accounted for by correcting just the norm bound, replacing $|v| \leq 1$ with
\begin{equation}
|v|\le F_\lambda[v]\,,
\end{equation}
leaving both the divergencelessness condition and the objective untouched. Here $F_\lambda$ can depend on both $v$ and its derivatives.

The constant $\lambda$ will always be considered to be perturbatively small, allowing both the flow $v^\mu$ and the norm bound to be expressed as perturbative expansions. Thus we will write
\begin{equation}
F_\lambda[v]=1+\sum_{n=1}^\infty\lambda^nf_n[v]\,.
\end{equation}
We will compute the functions $f_1$ and $f_2$. Although the entropy functional \eqref{eq:RT formula2} has no order-$\lambda^2$ term, it turns out that such a term is necessary in the norm bound. The reason is that maximizing the flux subject to a norm bound with a first-order correction leads to a second-order term in the flux, which must then be cancelled with an explicit second-order correction to the norm bound. Our general result can be found in subsection \ref{sec:combined}, and its application to Gauss-Bonnet gravity in Eq.\ \eqref{GB final}.

We derive our results by two different methods. The first, which we call the bottleneck method, is described in section \ref{sec:Bottleneck method}. It asks what the norm bound needs to be such that the maximum flux of a vector field defined on a given surface $m$ equals $a_\lambda(m)$. This implies that, for any $m\sim A$, $a_\lambda(m)$ bounds the flux through $A$ of any divergenceless vector field. The tightest bound is $\int_Av\le\min\limits_{m\sim A}a_\lambda(m)$. We then ask whether this inequality can be saturated, in other words whether a vector field defined on the minimizing surface and satisfying the norm bound can be extended to one that is defined everywhere in the space while respecting the divergenceless condition and the norm bound. This issue is non-trivial, and discussed in detail. The bottleneck method is general enough that it can in principle be used to write any higher-curvature HEE prescription in an equivalent bit-thread formulation. We apply the bottleneck method to the specific case of Gauss-Bonnet (GB) gravity and discuss how to incorporate the accompanying Gibbons-Hawking-York (GHY) boundary term. The second method, described in section \ref{sec:Lagrange dualization}, applies the machinery of Lagrange dualization of convex programs to the corrected min cut problem. This method is more straightforward, but it requires the corrected min cut problem to define a convex program, which is only true under restrictive assumptions, either when the minimal area surface has a high degree of symmetry or $\TCA$ has a particularly simple form. The bottleneck method on the other hand is always valid. For GB HEE both methods are valid when considering boundaryless bulk entangling surfaces without extrinsic curvature. Despite the limited applicability of Lagrange dualization to our problem, we include it because it is a non-trivial self-consistency check in results. In section \ref{sec: threadpaths}, we consider cases where the norm bound is a function purely of flow direction from the viewpoint of bit threads.

\section{Bottleneck method}
\label{sec:Bottleneck method}

In this section we explain the bottleneck method and use it to derive the norm bound corrections.

\subsection{Method}

We will retain, from the max flow-min cut theorem, the objective $\int_A v$ as well as the divergenceless constraint $\nabla_\mu v^\mu=0$. As a result, we still have, for any surface $m$ homologous to $A$,
\begin{equation}
\int_mv=\int_Av\,,
\end{equation}
so we can measure the flux through any $m$. Let $m_\lambda^*$ be the surface that minimizes the functional $a_\lambda$ whose minimum gives $S(A)$ in the corrected RT formula \eqref{eq:RT formula2}. We wish to find a norm bound
\begin{equation}\label{eq:normbound}
|v|\le F_\lambda
\end{equation}
such that the maximum flux equals
\begin{equation}\label{eq:saturation1}
\max_v\int_{m}v = a_\lambda(m^*_\lambda) =\int_{m^*_\lambda} \sqrt{g}\left(1+ \lambda \tilde {\mathcal A} \right).
\end{equation}
with $\lambda \tilde{\mathcal{A}}$ the perturbative correction to the HEE area functional. This is not a trivial task because $F_\lambda$ is not allowed to depend explicitly on $m_\lambda^*$, only on $v$ and the local geometry since the norm bound \eqref{eq:normbound} will be imposed everywhere without reference to any particular surface.

Let us fix a surface $m$ homologous to $A$, and further restrict $m$ to be perturbatively close to the $\lambda=0$ RT surface $m_0^*$. To see that this restriction does not exclude any potential bottlenecks, note that corrections to the flow must be perturbatively small, so the position where the bit threads are maximally packed can only move a perturbatively small distance, assuming no flat directions in the position of the RT surface. A useful property of minimal area surfaces such as $m_0^*$ is that $K$ the trace of the extrinsic curvature vanishes. This implies, since $m$ is perturbatively close to $m_0^*$,  
\bne K = \mathcal{O}(\lambda) \text{ on $m$.} \ene  
We will do a local analysis on $m$, thus in this subsection the vector field $v$ is defined only on $m$ and its neighborhood. Only when it is established that the norm bound we derive has flow solutions that do not violate it in the neighborhood of $m$, will $v$ be extended off the surface onto the entire time slice. We will derive a norm bound of the form \eqref{eq:normbound} such that
\bne \label{eq:bottleneck_method}
\max_{v}\int_mv = a_\lambda(m)\,,\quad |v|\le F_\lambda[ v]\text{ on $m$}\,.
\ene
Separating $v$ into its norm $|v|$ and direction $\hat v$, the flux is
\begin{equation} \label{eq:flux through arbitrary surface}
\int_mv = \int_m\sqrt{\tilde g}\,u_\mu v^\mu = \int_m\sqrt{\tilde g}\,u_\mu\hat v^\mu\,|v|\,
\end{equation}
with $u$ the unit normal on $m$.
To maximize the flux for a given direction field $\hat v$ (assuming $u_\mu\hat v^\mu\ge0$, as it will be) corresponding to a fixed orientation of the bit threads on $m$, the threads should clearly be maximally packed, saturating the norm bound such that
\begin{equation}\label{eq:saturation}
|v|=F_\lambda[v]\quad\text{on }m\,,
\end{equation}
so what we want to maximize and match is
\bne \label{eq:bottleneck_method2}
\max_{v} \int_m\sqrt{\tilde g}\, u_\mu \hat v^\mu F_\lambda[ v]= \int_m \sqrt{\tilde{g}}(1+\lambda \TCA ).
\ene

In subsection \ref{sec:obstruction}, we will find the conditions on the maximizing vector field and the surface $m$ such that the vector field can be extended off of that surface while respecting both the divergencelessness constraint and the norm bound. In subsection \ref{sec:corrections to the maximum bit thread density} we derive some useful perturbative expansions in $\lambda$, and in \ref{sec:Order by order evaluation} we will find a suitable $F_\lambda$ by first considering the vector field $v$ on $m$, and requiring that \eqref{eq:saturation1} hold. 
This establishes 
\bne
\max_v \int_A v = a_\lambda (m_\lambda^*)
\ene
as the maximum flux from $A$ is bounded by the maximum flux through the bottleneck, which when 
\eqref{eq:saturation1} holds is also the surface that minimizes $a_\lambda (m)$.

In this section Gaussian normal coordinates (GNC) will sometimes be used, with the notation 
\bne ds^2 = dz^2 + \tilde g_{ij} dx^i dx^j. \ene
Thus $v^z$ is the component of $v$ normal to hypersurfaces of constant $z$, and $v^i$ the tangential components. We take $m$ to be the surface with $z = 0$. 
\subsection{Obstruction equations} \label{sec:obstruction}
The flux $\int_A v$ through the boundary subregion $A$ is equal to the flux $\int_m v$ through any surface $m$ homologous to $A$, from the divergenceless condition on $v$ and Stoke's theorem. This is assuming there is no obstruction to the flow, that is, given some $v$ on $m$, it is possible to extend $v$ from $m$ to the boundary ($A \cup A^c$) without anywhere violating the norm bound. For $\lambda = 0$, this is true by the MFMC theorem. We would like to know whether for $\lambda \neq 0$ this holds true, or whether the maximally packed threads will inevitably collide.

Away from the bottlenecks the bit threads are far from maximally packed; stopping them from running into each other and violating the norm bound is simple, as there is a lot of space for maneuvering. On the minimal surface however, the threads are maximally packed and their directionality fixed; there is no choice on how to orient the threads in order to stop them colliding, so just off the surface is the most likely place for the norm bound to be violated. 

Let us calculate the necessary condition for the threads not to collide off a surface $m$, given that the norm bound on $v$ is saturated on that surface 
\bne |v|-F_\lambda [v] =0.\ene
If the difference between the norm bound and $F_\lambda [v]$ is anywhere positive, on or off the surface, then the norm bound has been violated. Assuming the flow to be smooth, the norm bound must be saturated to linear order in distance from $m$ 
\begin{equation} \label{eq:obstruction equation starting point}
\del_z (|v| - F_\lambda [v])|_{m} = 0,
\end{equation}
which may be written as
\bne \label{eq:linear obstruction}
v_\mu \nabla_z v^\mu =  F_\lambda [v] \nabla_z F_\lambda [v] .
\ene
To quadratic order in distance from $m$ the non-violation of the norm bound is expressed as a bound on the second normal derivative
\bne 
\nabla_z^2 (|v| - F_\lambda [v])|_m \leq 0.
\ene
which may be written as
\bne \label{eq:quadratic obstruction} v_\mu \nabla_z^2 v^\mu + \nabla_z v_\mu \nabla_z v^\mu - F_\lambda [v] \nabla_z^2  F_\lambda [v] - (\nabla_z F_\lambda [v])^2 \leq 0. \ene

\subsection{Perturbative expansions of the flow}\label{sec:corrections to the maximum bit thread density}
Here we introduce our notation for keeping track of $\lambda$ dependence, and derive useful perturbative expansions of the flow. The bulk geometry is a solution to the Einstein's field equations with higher curvature corrections, so has dependence on $\lambda$. The flow solution which maximizes the flux also depends on $\lambda$. In perturbative expansions care must be taken to keep track of the perturbative order of every term. The norm bound is expanded
\bne \label{eq:norm bound perturbative solution}
F_\lambda[v] = 1+ \sum^\infty_{n=1} \lambda^n f_{n}[v]
\ene
and the flow
\bne \label{eq:v_pert_sol}
v = v_0 + \sum^\infty_{n=1} \lambda^n v_n\,.
\ene
While $v_n$ has no $\lambda$ dependence, $f_n [v]$ does. For other quantities, perturbative expansions in $\lambda$ use $(n)$ to denote the nth order in the expansion. In our notation, by definition a quantity with a $(n)$ superscript never has $\lambda$ dependence.  For example, the metric we expand as
\bne g_{\mu\nu} = \sum^\infty_{n=0}\lambda^n g_{\mu\nu}^{(n)}. \ene
In GNC the $zz$ component of the metric is exactly 1 by definition,
\bne g_{zz}^{(0)} = 1, \qquad g_{zz}^{(n>0)} = 0\ene
while the tangential components do have $\lambda$ dependence with
\bne g_{ij}^{(n)} = \tilde{g}_{ij}^{(n)}. \ene
$\tilde{g}$ is the induced metric on $m$. In our notation the only quantities without a $(n)$ superscript which do \emph{not} have $\lambda$ dependence are $v_n$ and quantities which we have explicitly shown and stated to have no $\lambda$ dependence.

A useful simplification is made using a result from the zeroth-order RT bit threads, that the maximizing flow on the minimal area surface equals the unit normal, $v_0 |_{m_0^*} = u$. As $m_0^*$ and $m$ are perturbatively close, $v$ will still be normal to $m$ at zeroth-order and so
\bne v_0|_m = u, \ene which in turn implies that $|v_0|$ has no $\lambda$ dependence as
\bne \begin{split} |v_0| &= (g^{(0)}_{\mu\nu}v_0^\mu v_0^\nu + \lambda g^{(1)}_{\mu\nu}v_0^\mu v_0^\nu + ...)^{1/2} \\
&= (g^{(0)}_{zz} + \lambda g^{(1)}_{zz}+ ...)^{1/2}\\
&= 1 \end{split} \ene
so that on $m$
\bne |v_0|= |v_0|^{(0)} =1,\qquad |v_0|^{(n>0)} = 0. \ene
$v_0$ and $|v_0|$ do not have $\lambda$ dependence, so neither does $\hat{v}_0$ or the projection tensor
\bne \begin{split} {P^{\mu}}_\nu [\hat v_0] &:= {\delta^{\mu}}_\nu - \hat v_0^\mu \hat v_{0\nu} \\
&= {\delta^{\mu}}_\nu - g_{\nu\rho} \hat v_0^\mu \hat v^\rho_{0}\\
&= {\delta^{\mu}}_\nu - {\delta^z}_\nu \hat v_0^\mu 
. \end{split} \ene

Taking the norm of $v$ and perturbatively expanding gives
\bne \label{eq:vnorm} \begin{split}
|v| &= |v_0| + \lambda \hat v_{0\mu}v_1^\mu + \frac{\lambda^2}{|v_0|} ( v_{0\mu} v_2^\mu + \frac{1}{2} v_{1\mu}v_1^\mu - \frac{1}{2}(v_{0\mu}v_1^\mu)^2)+ \op(\lambda^3)\\
&= 1+ \lambda v_1^z + \lambda^2 (v_2^z + \frac{1}{2} v_{1i} v_1^i )+ \op(\lambda^3)\\
\end{split} \ene
When $|v|$ saturates its norm bound, we can use the results derived so far to perturbatively expand both sides of \eqref{eq:saturation}, finding
\bne \label{eq:v1z norm bound saturation}  v_1^z = (f_1 [v])^{(0)} \ene
and 
\bne v_2^z + \frac{1}{2}\tilde{g}^{(0)}_{ij} v_1^i v_1^j = (f_1 [v])^{(1)} + (f_2 [v])^{(0)}. \ene
The direction of $v$ can be expanded in $v_n$ to
\bne \label{eq:v hat lambda expansion} \begin{split}
\hat{v}^\mu &= 
\hat{v}_0^\mu + \lambda \frac{{P^\mu}_\nu [\hat{v}_0] v_{1}^{\nu}}{|v_0|} - \lambda^2 \frac{(\hat{v}_0^\mu {P^\rho}_\nu [\hat{v}_0] + 2 \hat{v}_0^\rho {P^\mu}_\nu [\hat{v}_0]) v_1^\nu v_{1\rho}-2|v_0| {P^\mu}_\nu [\hat{v}_0] v_2^\nu }{2|v_0|^2} + \op(\lambda^3) \end{split} \ene
We only need the normal component of $\hat v$ for our procedure; see equation \eqref{eq:bottleneck_method2}. On $m$ this is
\bne \begin{split} \label{eq:v hat z}
\hat{v}^z &= 1 - \lambda^2 \frac{v_{1i} v_1^i}{2} + \op(\lambda^3) \,.
\end{split} \ene
Before proceeding with the maximization of the flux order by order in $\lambda$, note that while $\sqrt{\tilde{g}}$ has a $\lambda$ expansion it is common to both the area functional and flux sides of \eqref{eq:bottleneck_method2} and so is a spectator; while not left out, it will be ignored. 
\subsection{Maximization of flux} \label{sec:Order by order evaluation}
In this subsection we maximize the flux order by order in $\lambda$, making use of the perturbative expansions \eqref{eq:vnorm} and \eqref{eq:v hat z}, and evaluate the obstruction equations \eqref{eq:linear obstruction} and \eqref{eq:quadratic obstruction}, in order to determine $F_\lambda [v]$. At each order in $\lambda$, there are three pieces of information that can be used to constrain $v$ on and off $m$,
\begin{enumerate}
\item The norm bound $|v|$ is saturated on $m$.
\item The direction $\hat{v}$ is such that the flux through $m$ is maximized.
\item The norm bound cannot be violated anywhere off $m$.
\end{enumerate}
Information that can be found about $f_n$ or $v_n$ at a given order in $\lambda$ can be used at higher-order.  
\sss{Zeroth order}
\emph{Norm bound:} The norm bound to zeroth-order is
\bne \label{eq:zeroth-order norm bound} |v|^{(0)} = (\tilde{g}_{ij}^{(0)} v_0^i v_0^j + (v_0^z)^2)^{1/2} = 1. \ene 
\emph{Flux:} The flux to zeroth order in $\lambda$ is 
\bne 
\int_m \sqrt{\tilde g} \hat{v}_0^z
\ene
which given the zeroth order norm bound \eqref{eq:zeroth-order norm bound} is maximized when 
\bne \label{eq:v0 on m} 
v_0^z = 1, \qquad v_0^i = 0
\ene
As stated earlier this is a known result from RT bit threads, derived here using the novel bottleneck method.
 
That $v_0 = u$ on $m$ allows us to replace $v_0$ with $u$ in functionals which do not contain derivatives perpendicular to the surface, for example $k_{ij}[v_0] = K_{ij}$. Those normal derivatives are thus far unconstrained, for example $v_{0}^i$ on $m$ is known, while $\del_z v_0^i$ is not.

\emph{Linear obstruction equation:} The zeroth-order of the linear obstruction equation is
\bne (v_\mu \nabla_z v^\mu)^{(0)} = \nabla_z v_0^z =  \del_z v_0^z  = 0. \ene
From this we can show that the trace of the extrinsic curvature on $m$ must vanish to zeroth-order, using
\bne \label{eq:zeroth order linear obstruction}
 \del_z v_0^z = (\nabla_z v^z)^{(0)} =  (\nabla_\mu v^\mu -\nabla_i v^i )^{(0)} = - (\Gamma^i_{iz} )^{(0)}= - K^{(0)}. \ene
which implies
\bne K^{(0)} =0. \ene
That the zeroth-order component in the trace of the extrinsic curvature vanishes is thus a no-obstruction constraint on $m$. This is consistent with the restriction made earlier that $m$ be perturbatively close to a minimal area RT surface. For RT bit threads this is a known result, that it is not possible to extend bit threads off a surface on which the threads are maximally packed without violating the norm bound, unless it is a RT surface, with $K =0$.

\emph{Quadratic obstruction equation:} The zeroth order in the quadratic obstruction equation \eqref{eq:quadratic obstruction} is 
\bne \begin{split} \label{eq:zeroth order in the quadratic obstruction equation}
& (v_{0\mu} \nabla_z^2 v_0^\mu + \nabla_z v_{0\mu} \nabla_z v_0^\mu)^{(0)} \\
 &= (\nabla_z^2 v_0^z + \tilde{g}_{ij} \nabla_z v_{0}^i \nabla_z v_{0}^j)^{(0)} \\
 &= (R_{zz} - \nabla_i \nabla_z v_0^i + \del_z v_{0i} \del_z v_{0}^i)^{(0)}  \leq 0. 
 \end{split} \ene
where the last line follows from
\bne \nabla_z^2 v^z = \nabla_z(\nabla_\mu v^\mu - \nabla_i v^i) = -\nabla_z \nabla_i v^i = R_{\mu z}v^\mu - \nabla_i \nabla_z v^i. \ene
Note that in expressions of the form $(...)^{(0)}$, $v$ can be replaced with $v_0$ and vice versa. The bound \eqref{eq:zeroth order in the quadratic obstruction equation} is a constraint on how $v_0$ changes off the surface, and the max-flow-min-cut theorem states that for all minimal area surfaces with $K^{(0)}=0$ there is always an obstructionless flow $v_0$, and hence the above constraint inequality places no further condition on $m$ at this order. If at higher orders we need to maximize flux over $\nabla_z v_0^i$ then this inequality will be important, but we will see at second order how the dependence of flux on $\nabla_z v_0^i$ can be removed with a suitable choice for $f_2 [v]$.

\sss{First order}
\emph{Flux:} The flux to first order in $\lambda$ is 
\bne \left ( \int_m v \right )^{(1)}= \int_m \sqrt{\tilde{g}} ((\hat{v}^z)^{(0)}(F_\lambda [v])^{(1)} + (\hat{v}^z)^{(1)}(F_\lambda [v])^{(0)} = \int_m \sqrt{\tilde{g}} (f_1[v_0])^{(0)} \ene
Comparing this to the first-order term in $a_\lambda(m)$ implies
\bne\label{eq:firstorder}
(f_1 [v_0])^{(0)} = \tilde{\mathcal{A}}^{(0)}. 
\ene
\emph{Norm bound:} The norm bound saturation \eqref{eq:v1z norm bound saturation} on $m$ to first order is
\bne v_1^z = (f_1 [v_0])^{(0)} = \tilde{\mathcal{A}}^{(0)}, \ene
however the tangential components of $v_1$ are undetermined at this order.

\emph{Linear obstruction equation:} The first-order in the linear obstruction equation \eqref{eq:linear obstruction} is 
\bne \begin{split} \label{eq:linear and first order obstruction} 
&(v_\mu \nabla_z v^\mu - F_\lambda [v] \nabla_z F_\lambda [v])^{(1)} \\
&=(\nabla_z v_1^z + v_{1i}\del_z v_0^i - \del_z f_1[v])^{(0)}\\
&= 0.
\end{split}\ene 
where we have used $\del_z v_0^z = 0$ on $m$.

\emph{Quadratic obstruction equation:} The first order of the quadratic obstruction equation does not have any impact on our flux maximization, more details are given in appendix \ref{sec:Linear order quadratic obstruction equation}.

\sss{Second order}
\emph{Flux:} The flux to second order in $\lambda$ is
\bne\label{eq:second-order} \begin{split}
\left(\int_m v\right)^{(2)} &=\int_m\sqrt{\tilde g} ( \hat{v}_z^{(0)}(F_\lambda [v])^{(2)} +\hat{v}_z^{(1)}(F_\lambda [v])^{(1)}+\hat{v}_z^{(2)}(F_\lambda [v])^{(0)}) \\
&=  \int_m\sqrt{\tilde g} \left(f_1[v]^{(1)} + f_2 [v]^{(0)} - \frac{\tilde{g}_{ij}^{(0)}v_1^i v_1^j}{2}\right)\\
&= \int_m\sqrt{\tilde g} \left((f_1 [v_0])^{(1)} +(f_1[v]-f_1[v_0])^{(1)}  + f_2 [v]^{(0)} - \frac{\tilde{g}_{ij}^{(0)}v_1^i v_1^j}{2}\right)\\
\end{split}\ene
We will maximize this contribution to the flux with respect to $v_1$, so it is important to know the $v_1$ dependence of each term, to this end in the last line we separated $(f_1 [v])^{(1)}$ into terms containing only $v_0$, and those exactly linear in $v_1$. 

Suppose we took the functional $\TCA [u]$ and replaced $u$ with $\hat{v}$. Let us call that functional $\tilde a [\hat v]$. As $\hat v_0 = u$ on $m$, and $\TCA$ contains only derivatives projected tangentially to the surface we have
\bne \label{eq:tilde a is a good choice} \tilde a [\hat v_0] = \TCA, \ene
If we choose $f_1$ to equal $\tilde a$ then we have
\bne \label{eq:all orders choice for f1} f_1 [v_0] = \tilde{\mathcal{A}} \ene
and the correction to the HEE surface functional is captured to all orders in $\lambda$. There is still work to do however as there are terms left over in the second order flux, which as everything in the HEE functional has been accounted for must equal zero. These additional terms come from the flow being perfectly normal to $m$, only to zeroth order in $\lambda$, the higher order corrections to the norm bound ($f_2$ and above) exist to cancel overcorrections to the flux. 

We would like to keep $f_1$ as general as possible, so note that we can add
\bne f_{1b} [v] = \sum_{n=1}^\infty p_n [v] (v^\mu \del_\mu |v|)^n \ene
to $f_1 [v]$ without changing $f_1 [v_0]$ on $m$, as $v_0^\mu \del_\mu |v_0| = \del_z v_0^z = 0$ on the surface and so \eqref{eq:all orders choice for f1} is still satisfied. $f_{1b}[v_0]$ does not therefore affect the flux, but $f_{1b}$ is important for the flow to be obstructionless. $p_n$ are unfixed functions. $v^\mu \del_\mu |v|$ measures change in bit thread number density tangential to the flow. The first order correction to the norm bound thus has two components, 
\bne f_1 [v] = \tilde a[\hat v] + f_{1b} [v], \ene
one which captures the surface functional correction $\tilde A$, and the other which ensures flow is obstructionless.

Let us return to the second order flux and calculate the contribution from $f_{1b}$,
\bne \begin{split} 
&\int_m \sqrt{\tilde g} (f_{1b}[v] - f_{1b}[v_0])^{(1)} \\
= &\sum_{n=1}^{\infty} \int_m \sqrt{\tilde g} (p_n [v] (v^\mu \del_\mu |v|)^n - p_n [v_0] (v_0^\mu \del_\mu |v_0|)^n )^{(1)}\\
= &\int_m \sqrt{\tilde g} (p_1 [v])^{(0)} (\del_z |v|)^{(1)} \\
= &\int_m \sqrt{\tilde g} (p_1 [v] \del_z f_{1} [v])^{(0)}.
\end{split} \ene 
To reach the last line we have used the first order linear obstruction constraint
\bne (\del_z (|v| - (1 + \lambda f_1[v]))^{(1)} = 0. \ene

Now let us calculate the contribution to the second order flux from $(\tilde a [v] - \tilde a [v_0])^{(1)}$. As we will be performing a functional variation around $v = v_0$,
we need to understand what derivative terms of $v$ can appear. As derivatives of $u$ in $\TCA$ must be projected tangential to $m$, all derivative terms of $\hat v$ in $\tilde a [\hat v]$ must be projected onto the normal subspace of $v$. A consequence of this is that terms involving the normal derivatives of $v$ vanish at zeroth order, for example, suppose that $\tilde{a} [\hat v]$ is the trace of the extrinsic curvature
\bne k [\hat v] = P^{\mu\nu} [\hat v] \nabla_\mu \hat v_\nu, \ene
then terms such as
\bne \left (\frac{\del \tilde k [\hat v]}{\del (\nabla_z \hat v_\mu)} \right )^{(0)} = (P^{z\nu} [\hat v])^{(0)}  = g^{z\nu} - u^z u^\nu = 0. \ene
This leaves just derivatives tangential to $m$, which can be integrated by parts to strip off all the derivatives acting on $v_1$, in a fashion similar to the derivation of the Euler-Lagrange equation. The contribution from $\tilde a$ is 
\bne \label{contribution from tilde a} \begin{split} 
&\int_m \sqrt{\tilde g} (\tilde a [\hat v] - \tilde a [\hat v_0])^{(1)}\\
&= \int_m \sqrt{\tilde g} \left(\tilde a \left[\hat v^\mu_0 + \lambda \frac{{P^{\mu}}_\nu [\hat v_0] v_1^\nu}{|v_0|}\right] - \tilde a [\hat v_0]\right)^{(1)}\\
&=\int_m \sqrt{\tilde g} \left(\frac{\del \tilde a [\hat v]}{\del \hat v^\mu} \frac{{P^{\mu}}_\nu [\hat v_0] v_1^\nu}{|v_0|} + \frac{\del \tilde a [\hat v]}{\del (\nabla_\rho \hat v^\mu )}\nabla_\rho \left(\frac{{P^{\mu}}_\nu [\hat v_0] v_1^\nu}{|v_0|}\right) +...\right)^{(0)} \\
&=\int_m \sqrt{\tilde g} \left(\frac{\del \tilde a [\hat v]}{\del \hat v^i} v_1^i + \frac{\del \tilde a [\hat v]}{\del (\nabla_j \hat v^\mu )}\nabla_j \left(\frac{{P^{\mu}}_\nu [\hat v_0] v_1^\nu}{|v_0|}\right) +...\right)^{(0)}\\
&=\int_m \sqrt{\tilde g} \left( \left( \frac{\del \tilde a [\hat v]}{\del \hat v^i}  -\nabla_j \frac{\del \tilde a [\hat v]}{\del (\nabla_j \hat v^i )}+ ... \right )v_1^i \right)^{(0)} + \text{boundary terms}\\
&=\int_m \sqrt{\tilde g} \left( \zeta_i [v] v_1^i \right)^{(0)} + \text{boundary terms}\\
 \end{split} \ene
 with the definition
 \bne \label{eq:zeta definition2} 
 \zeta_{i} [v] := \left( \frac{\del \tilde a [\hat v]}{\del \hat v^i}  -\nabla_j \frac{\del \tilde a [\hat v]}{\del (\nabla_j \hat v^i )}+ \nabla_k \nabla_j \frac{\del \tilde a [\hat v]}{\del (\nabla_j \nabla_k \hat v^i )} - ... \right )
\ene
Let us assume that the boundary terms vanish. We will explicitly show they do for GB HEE.

As $v_0 = \hat v_0$ on $m$, we could have replaced any of the $\hat v_0$ terms in $\tilde a [\hat v_0]$ with $v_0$. Then there would be additional terms in the second order flux involving $v_0$ and $v_1^z = \tilde{\mathcal{A}}^{(0)}$. As functions purely of $v_0$ such as these are easily removed with a suitable choice for $f_2$, as we will see, we are in effect only shuffling terms between $f_1$ and $f_2$ and nothing is lost by taking $\tilde a$ to be purely a function of the direction field $\hat v$.
Substituting both the contributions from $\tilde a$ and $f_{1b}$ into the second order flux gives
\bne\label{eq:second-order1} \begin{split}
\left(\int_m v \right)^{(2)} = \int_m\sqrt{\tilde g} \left(\TCA^{(1)} +  \zeta_i [v] v_1^i + p_1 [v]\del_z f_1 [v] + f_2 [v] - \frac{v_1^i v_{1i}}{2}\right)^{(0)}
\end{split}\ene
For now we will assume there is no constraint on $v_1^i$ from the obstruction equations, maximize the second order flux with respect to $v_1^i$, then find a $p_1 [v]$ such that the obstruction equation is satisfied for this maximizing value of $v_1^i$. The maximizing value of $v_1^i$ is 
\bne v_{1i} = (\zeta_i [v])^{(0)} \ene
for which the second order flux is 
\bne\label{eq:second-order2} \begin{split}
 \int_m\sqrt{\tilde g} \left(\TCA^{(1)} + p_1 [v]\del_z f_1 [v] + f_2 [v] + \frac{\zeta^i [v] \zeta_i [v]}{2}\right)^{(0)}.
\end{split}\ene
To determine $f_2$ we equate this with the second order of $a_\lambda (m)$, 
\bne (a_\lambda (m) )^{(2)} = \int_m \sqrt{ \tilde g} \TCA^{(1)} \ene
which implies
\bne \begin{split}
 (f_2 [v])^{(0)} &= \left(-\frac{1}{2}\zeta_i [v]\zeta^i [v] - p_1 [v] \del_z f_1 [v] )\right)^{(0)} \\
&= \left(-\frac{1}{2}\zeta_i [v]\zeta^i [v] - p_1 [v] (\del_z \tilde a [\hat v] + p_1 [v] \del_z^2 |v| )\right)^{(0)}. 
\end{split} \ene
While this does not fully determine $f_2$, since we will not continue to third-order in $\lambda$ any choice satisfying the above constraint is adequate for our purposes, so let us take the simplest, defined over the extended flow domain
\bne f_2 [v] = -\frac{1}{2}P_{\mu\nu} [\hat v] \zeta^\mu [v]\zeta^\nu [v] - p_1 [v] v^\mu \del_\mu \tilde a [v] - (p_1 [v])^2 v^\mu v^\nu \nabla_\mu \nabla_\nu |v| .\ene
To determine $p_1$, we return to the first order linear obstruction equation \eqref{eq:linear and first order obstruction},
\bne \begin{split} \label{eq:obstreq} &(\nabla_z v_1^z + v_{1i}\del_z v_0^i - \del_z f_1[v])^{(0)}\\
= &(-\nabla_i v_1^i - K^{(1)} + v_{1i}\del_z v_0^i - \del_z \tilde a [\hat v] - p_1 [v] \del_z^2 |v|)^{(0)} \\
= &0
.\end{split}\ene
where we have used the divergencelessness of $v$ to relate $\nabla_z v_1^z$ to derivatives tangential to $m$,
\bne \begin{split} 
(\nabla_\mu v^\mu)^{(1)} &= (\nabla_\mu v_0^\mu + \lambda \nabla_\mu v_1^\mu)^{(1)} \\
&= (\del_z v_0^z + \del_i v_0^i + \Gamma^i_{i\mu}v_0^\mu)^{(1)} + (\nabla_\mu v_1^\mu)^{(0)} \\
&= K^{(1)} + (\nabla_z v_1^z + \nabla_i v_1^i)^{(0)} \\
&= 0.
\end{split} \ene
Before we maximized the second order flux with respect to $v_1^i$, this obstruction equation \eqref{eq:obstreq} contained two unconstrained components of the flow $v_1^i$ and $\del_z v_0^i$, plus any further unconstrained derivatives of $v_0$ from one's choise of $p_1$. If we simply solve the obstruction equation \eqref{eq:obstreq} for $p_1$
\bne \begin{split} \label{eq:p1 zeroth order} ( p_1 [v] )^{(0)} &= \frac{(\del_z |v|)^{(1)} - (\del_z \tilde a [\hat v])^{(0)}}{(\del_z^2 |v|)^{(0)}} \\ 
&= \left ( \frac{1}{\del_z^2  |v|}( \zeta_i [v] \del_z v_0^i - \nabla_i \zeta^i [v] - \del_z \tilde a [v] - K^{(1)})\right )^{(0)}. \end{split} \ene
and define $p_1 [v]$ such that it evaluates to this on $m$, then no matter what values the unconstrained components of the flow take there is no obstruction to the flow, at first order in $\lambda$ and first order in distance from $m$. We were free to choose any function for $p_1$, however besides the special choice given above, \eqref{eq:obstreq} gives a constraint on $v_1^i$ in terms of $\del_z v_{0i}$ and whatever other unconstrained derivatives of $v_0$ appear in the choice for $(p_1[v])^{(0)}$, and this constraint needs to be imposed when maximizing the second order flux with respect to $v_1^i$. The above choice for $p_1$ is merely the most convenient.

The choice for $p_1$ given by \eqref{eq:p1 zeroth order} is singular whenever $\del_z^2 |v_0| = 0$ and the numerator is non-zero. Away from $m_\lambda^*$ this is not an issue as regions where the flow capacity is infinite do not affect the bottleneck position. As the flow always seeks to maximize flux, we only need to assume the existence of any $v_0$ for which $\del_z^2 |v_0| \neq 0$ everywhere (corresponding to threads always moving apart), or even if no such $v_0$ exists, that there is not a new bottleneck created. We also need to argue that the flow can not take advantage of this choice of $p_1$ in order to increase the capacity of the bottleneck. The correction to the norm bound on $m$ from $p_1$ is
$\lambda (p_1 [v] \del_z |v|)^{(0)}$. Now $(\del_z |v|)^{(0)}$ always equals $0$ on $m$, however the flow can still try to increase capacity by choosing $\del_z^2 |v_0| =0$ in which case we need to apply L'H\^{o}pital's rule to evaluate the ratio $\del_z |v_0| / \del^2_z |v_0|$. $\del_z^3 |v_0|$ must be zero on $m$ for there to be no obstruction, so we consider $\del_z^3 |v_0| / \del_z^4 |v_0|$. Again the flow can take $\del_z^4 |v_0| =0$, and so on, for the ratio $\del_z^{2n-1} |v_0| / \del_z^{2n} |v_0|$, the limit of which is where $|v_0| = 1$ everywhere.
 
Now that we have found a condition such that there is no obstruction to the flow, we may extend $v$ off the surface such that it is defined throughout the time slice. Let us choose a function for $p_1$ whose domain is over this extended flow, which when evaluated on $m$ satisfies the constraint \eqref{eq:p1 zeroth order},
\bne p_1 [v] =\frac{1}{v^\mu v^\nu \nabla_\mu\nabla_\nu |v|} ({P^\mu}_\nu [v] (\zeta_\mu [v] v^\rho \nabla_\rho v^\nu - \nabla_\mu \zeta^\nu [v]) - v^\rho \nabla_\rho \tilde a [v] - k_1 [v]). \ene  
$k_1 [v]$ is the function formed by replacing $u$ in $K^{(1)}$ with $\hat v$. This definition for $p_1 [v]$ makes no reference to any particular surface, $K^{(1)}$ and hence $k_1 [v]$ can be derived directly from the HEE surface functional by taking a variational derivative, only the equation of motion of the minimizing surface is needed, not its solution. 

There are terms in $f_{1b} [v]$ that are higher order in $v^\mu \del_\mu |v|$ that are still unfixed at this order in $\lambda$, however as we will not proceed to the next order we are free to set them to zero, $p_{(n\geq 2)} [v] = 0$, giving
\bne f_1 [v] = \tilde a [\hat v] + p_1 [v] v^\mu \del_\mu |v|. \ene

\emph{Norm bound:} We can determine the value of $v_2^z$ on $m$ from the saturation of the norm bound to second-order. Using \eqref{eq:vnorm} and \eqref{eq:v hat z},
\bne \begin{split} \int_m (|v| \hat v^z )^{(2)} &= \int_m (|v|^{(2)} + (\hat v^z )^{(2)}) \\
&= \int_m  v_2^z  \\
&= \int_m \tilde{\mathcal{A}}^{(1)}.
\end{split} \ene
This gives the component of $v_2$ normal to $m$,
\bne  v_2^z =  \tilde{\mathcal{A}}^{(1)}\ene

\emph{Obstruction equations:} The second-order of the obstruction equations gives conditions on $v_2^i$ which would only be relevant if we continued to maximizing flux at third-order.

\subsection{Combined results}
\label{sec:combined}

Combining all results from zeroth to second-order, the maximizing value of $v$ on $m$ is
\bne \label{eq:v_1 on m} v^\mu|_m = (1,\vec{0}) + \lambda ( \tilde{\mathcal{A}}^{(0)} , \zeta^i [v_0]^{(0)}) + \lambda^2 (\tilde{\mathcal{A}}^{(1)}, v_2^i) + \op (\lambda^2 ). \ene 
with $v_2^i$ unknown at second order in $\lambda$.  
The norm bound extended off of $m$, without reference to any surface, is
\bne |v| \leq 1 + \lambda (\tilde{a}[\hat v] + p_1 [v] v^\mu \del_\mu |v|) - \lambda^2 \left(\frac{1}{2}P_{\mu\nu} [\hat v] \zeta^\mu [v]\zeta^\nu [v] + p_1 [v] v^\mu (\del_\mu \tilde a [v] + p_1 [v]  v^\nu \del_\mu \del_\nu |v| )\right)+ \op (\lambda^3) \ene
with $p_1$ defined as
\bne p_1 [v] :=\frac{1}{v^\mu v^\nu \nabla_\mu \del_\nu |v|} ({P^\mu}_\nu [v] (\zeta_\mu [v] v^\rho \nabla_\rho v^\nu - \nabla_\mu \zeta^\nu [v]) - v^\rho \del_\rho \tilde a [\hat v] - k_1 [v]). \ene
and $\zeta$ defined as 
\bne \label{eq:zeta definition} 
 \zeta_{\mu} [v] := \left( \frac{\del \tilde a [\hat v]}{\del \hat v^\mu} - P^{\nu\rho}[\hat v]\nabla_\nu \frac{\del \tilde a [\hat v]}{\del (\nabla^\rho \hat v^\mu )}+ P^{\nu\rho}[\hat v] P^{\sigma\omega}[\hat v] \nabla_\nu \nabla_\sigma \frac{\del \tilde a [\hat v]}{\del (\nabla^\omega \nabla^\rho \hat v^\mu )} - ... \right ).
\ene
This is as high in orders of $\lambda$ as we will go. In principle one could continue the procedure of maximizing the flux and equating it to the HEE functional to even higher order, and this would continue to give corrections to the value of the flow on $m$ and the norm bound. At each order in $\lambda$ a new degree of freedom $v_n^\mu$ is added over which the flux is maximized, and corrections to the norm bound are added to correct for over/undershooting.

\subsection{Application to Gauss-Bonnet gravity}\label{sec:GB bottleneck}

Let us apply our results to Gauss-Bonnet (GB) gravity, where the correction to the surface functional is 
\bne \TCA = \tilde{R}. \ene
with $\tilde R$ the induced scalar curvature of the surface
\bne \tilde{R} = R - 2R_{\mu\nu} u^\mu u^\nu + ({K_\mu}^\mu)^2 - K_{\mu\nu}K^{\mu\nu}\ene 
and $K_{\mu\nu}$ the extrinsic curvature tensor
\bne
K_{\mu\nu}={P_\mu}^\rho[u]\nabla_\rho u_\nu .
\ene
Gauss-Bonnet gravity is the simplest extension to Einstein gravity that is a Lovelock theory. The Lagrangian in a Lovelock theory is a sum of Euler densities, 
\bne
\mathcal{L}_{2p} \coloneqq \frac{1}{2^{p}} \delta^{\nu_{1} \dots \nu_{2p}}_{\mu_{1} \dots \mu_{2p}}R^{\mu_{1}\mu_{2}}_{\nu_{1} \nu_{2}} \dots R^{\mu_{2p-1}\mu_{2p}}_{\nu_{2p-1} \nu_{2p}}\,,
\ene
quantities whose integrals are topological invariants in $2p$ dimensions. The equations of motion of such a theory contain only second derivatives of the metric, meaning that they require the same initial data as Einstein gravity. GB gravity includes, in addition to the usual cosmological constant ($p=0$) and Einstein-Hilbert ($p=1$) terms, the $p=2$ term:
\begin{equation}\label{eq:action}
I = \frac1{16\pi G_{\rm N}}\int\sqrt g\left(-2\Lambda+R+\lambda(R^2-4R_{\mu\nu}R^{\mu\nu}+R_{\mu\nu\lambda\sigma}R^{\mu\nu\lambda\sigma})\right)+\text{boundary terms}\,,
\end{equation}
where $\lambda$ is a parameter with dimensions of length-squared. For HEE in GB gravity, the entropy is given by minimizing a functional which includes the area plus the integrated induced Ricci scalar \cite{Hung:2011xb,deBoer:2011wk,Haehl2017}. The GB HEE functional is 
\begin{equation}\label{eq:adef2}
a_\lambda(m) \coloneqq
\int_{m} \sqrt{\tilde g}(1+ \lambda \tilde{R}) + 2\lambda \int_{\del m}\sqrt{\tilde h} \tilde K\,,
\end{equation}
and where we use tildes to denote quantites defined with respect to the induced metric $\tilde g_{ij}$ on the surface $m$, $\tilde K$ is the trace of the extrinsic curvature not of $m$ but $\del m$. The Gibbons-Hawking-York (GHY) boundary term in \eqref{eq:adef2} is necessary to give a well-posed variational problem. We should again emphasize that we could have chosen any higher-curvature correction to Einstein gravity to illustrate our method, as long as the entropy is given by minimizing a local functional on surfaces in the homology class of $A$. We will not be using any special properties of Lovelock theories, Gauss-Bonnet gravity is merely a simple extension to consider.

Before proceeding, we note an important caveat regarding the GB HEE formula. Naively, it gives $-\infty$ for the entropy of any region. This can be easily seen in $3+1$ bulk dimensions, where the surface $m$ is 2-dimensional and the $\lambda$ terms in \eqref{eq:adef2} are proportional to its Euler character $\chi(m)$:
\bne \label{eq:Gauss-Bonnet theorem}
\int_m\sqrt{\tilde g} \tilde{R} + 2\int_{\del m}\sqrt{\tilde h}\tilde K = 4\pi \chi(m) \,.
\ene
By adding small handles or spheres to the surface $m$, its Euler character can be made arbitrarily negative or positive without significantly changing the total area. Hence, for either sign of $\lambda$, the GB HEE formula, taken at face value, tells us the entropy will always be $-\infty$! However, one should remember that \eqref{eq:action} should be treated as an effective action, with $\lambda$ treated as a perturbative parameter, rather than assigned a finite value. Correspondingly, $\lambda$ should be treated as a perturbative parameter. In other words, the embedding coordinates of the surface $m$ should be written as a power series in $\lambda$, and then the surface functional minimized order by order in $\lambda$. 
In turn, all calculations demonstrating our methods on GB gravity will be done perturbatively in the Gauss-Bonnet parameter $\lambda$. 

By varying $a_\lambda (m)$ we find the equation of motion for $m_\lambda^*$, 
\bne (1+ \lambda \tilde{R})K - 2\lambda \tilde{R}_{ij}K^{ij}=0, \ene
which implies
\bne\label{eq:K GB} K^{(0)} = 0 ,\qquad K^{(1)} = 2 (\tilde{R}_{ij}K^{ij})^{(0)}. \ene 

The correction $\lambda \TCA$ cannot affect the zeroth order result, so we start at first-order. With $\TCA = \tilde R$, we have $\tilde a = r$ where
\bne \label{eq:definition of r}
\begin{split}
r[\hat v] &:= R - 2R_{\mu\nu} \hat v^\mu \hat v^\nu + ({k_\mu}^\mu[\hat v])^2 - k_{\mu\nu}[\hat v]k^{\mu\nu}[\hat v]\\
k_{\mu\nu}[\hat v] &:={P_\mu}^\rho[\hat v]\nabla_\rho \hat v_\nu\, ,\quad {P_\mu}^\nu[\hat v] := {\delta_\mu}^\nu - \hat v_\mu \hat v^\nu .
\end{split}
\ene
The leading order correction to the thread thickness is $\lambda r[v_0] = \lambda \tilde{R}$, so loosely speaking the more curved the surfaces which are perpendicular to the flow are, the more the thread thickness is affected, thicker or thinner depending on the sign of $\lambda$.

Given $\tilde a$, we next calculate the terms in $\zeta_i$,
\bne \pd{r [\hat v]}{\hat v^i} = -4R_{iz} + 2 {k_i}^j [\hat v] \nabla_z \hat v _j ,\qquad \nabla_j \pd{r}{(\nabla_j \hat v^i)} = - 2 \nabla_j {k^j}_i [\hat v] \ene
giving
\bne \label{eq:zeta in GB} \begin{split}(\zeta_i [v])^{(0)} &= (-4R_{iz} + 2 {k_i}^j [\hat v] \nabla_z \hat v _j + 2 \nabla_j {k^j}_i [\hat v])^{(0)}\\
&= 2({K_i}^j \nabla_z \hat v _j - R_{iz}) ^{(0)}
\end{split} \ene
which uses the identity
\bne \begin{split}
&R_{iz} = \nabla_j {K^j}_i - \del_i K \\
\implies &(R_{iz})^{(0)} = (\nabla_j {k^j}_i [\hat v])^{(0)}.
\end{split} \ene
In the derivation for the general case, we neglected the boundary terms arising from the integration by parts. For Gauss-Bonnet these are
\bne \int_{\del m} \sqrt{\tilde{h}} (k_{ij}[\hat v_0] \tilde{n}^i  v_1^j))^{(0)} \ene
In an asymptotically AdS spacetime, with spatial metric $ds^2\sim z^{-2}(dx^\mu)^2$ and cutoff $z=z_0$, $m_0^*$ has extrinsic curvature components $K_{ij}$ which remain finite on the boundary, while $\tilde n^i$ goes like $z_0$, so $(K_{ij}\tilde n^i)^2$ goes like $z_0^4$, and therefore vanishes as $z_0\to0$.

With $\zeta$ we can calculate $f_1$. Recall that
\bne f_1 [v] = \tilde{a} [\hat v] + p_1 [v] v^\mu \del_\mu |v| \ene
with $\tilde a [\hat v] = r [\hat v]$ for GB HEE, and that on $m$ $p_1$ is
\bne \left ( p_1 [v]  = \frac{1}{\del^2_z |v|}( \zeta_i [v] \del_z v_0^i - \nabla_i \zeta^i [v] - \del_z \tilde a [v] - K^{(1)})\right )^{(0)} \ene
Let us calculate each of the terms on the right individually, 
\bne \begin{split}(\zeta^i [v] \del_z v_i)^{(0)} &= 2(({k^i}_j [\hat v]\del_z \hat v^j - {R^i}_z)\del_z v_i )^{(0)} \\
 (-\nabla_i \zeta^i [\hat v])^{(0)} &= 2 (-\nabla_i {k^i}_j [\hat v] \del_z \hat v^j - {k^i}_j [\hat v] \nabla_i \nabla_z \hat v^j + \nabla_i {R^i}_z )^{(0)}\\ 
&= 2 (-R^i_z \del_z \hat v_i - {k^i}_j [\hat v] \nabla_i \nabla_z \hat v^j + \nabla_i {R^i}_z )^{(0)}\\ 
- K^{(1)} &= -2 (K^{ij}\tilde{R}_{ij})^{(0)} \\
(-\del_{z}r[v] )^{(0)}
&= (-\del_z R + 2\del_{z}R_{zz} + 4R_{zz}\del_{z}v^{z} + 4{R^i}_z\del_{z}v_{i}+2K^{ij}\nabla_{z}k_{ij}[v]-2{K^i}_i \nabla_{z}{k^j}_j[\hat v])^{(0)}\\
&= (-\del_z R + 2\del_{z}R_{zz} + 4{R^i}_z \del_z v_{i} +2K^{ij}\nabla_{z}k_{ij}[\hat v])^{(0)}\\
&= (-2\nabla_{i}{R^{i}}_{z} + 4{R^i}_z \del_z v_{i}  +2K^{ij}\nabla_{z}k_{ij}[\hat v])^{(0)} \\
 &= (-2\nabla_{i}{R^{i}}_{z} + 4{R^i}_z \del_z v_{i} + 2K^{ij}\nabla_{z}((\delta^{\rho}_{i}-v_{i}v^{\rho})\nabla_{\rho}v_{j}))^{(0)} \\ 
 &= (-2\nabla_{i}{R^{i}}_{z} + 4{R^i}_z \del_z v_{i} + 2K^{ij}(\nabla_{z}\nabla_{i}v_{j}-\del_z v_{i}\del_z v_{j}))^{(0)}\\
 &= (-2\nabla_{i}{R^{i}}_{z} +4{R^{i}}_{z}\del_z v_{i} + 2 K^{ij} (\nabla_i  \nabla_z v_{j} - \del_z v_{i} \del_z v_{j} + R_{jzzi}))^{(0)}.
\end{split} \ene
For $\del_z r$, in the first line the third and last terms vanish using $\del_z v_0^z = 0$ and $K^{(0)} = 0$, the third line makes use of the contracted Bianchi identity
\bne \del_\rho R = 2 \nabla_\mu {R^\mu}_\rho, \ene
and the final line uses the relation between the commutator of covariant derivatives and the Riemann tensor
\bne [\nabla_\mu , \nabla_\nu] V_\rho = R_{\rho\sigma\mu\nu}V^\sigma. \ene

Combining these contributions many terms cancel giving
\bne \begin{split}(p_1 [v])^{(0)} &=\left ( \frac{1}{\del_z^2 |v|}  2K^{ij}(\tilde{R}_{ij} - R_{jzzi}) \right )^{(0)}\\
&= \left ( \frac{1}{\del^2_z |v|}  2K^{ij}(R_{ij} - {K_i}^m K_{mj})\right )^{(0)}
\end{split}\ene
on $m$, using the GNC identities
\bne \tilde{R}_{ij} = R_{ij} + \del_z K_{ij} + K K_{ij} - 2 {K_{i}}^m K_{mj} \ene
and 
\bne R_{izzj} = \del_z K_{ij} - {K_i}^m K_{mj}.\ene
Thus for GB HEE, the norm bound defined without reference to any surface is
\begin{multline}\label{GB final}
|v| \leq 1 + \lambda (r[\hat v] + p_1 [v] v^\mu \del_\mu |v|) \\{}- \lambda^2 \left(\frac{1}{2}P_{\mu\nu} [\hat v] \zeta^\mu [v]\zeta^\nu [v] + p_1 [v] v^\mu (\del_\mu r [v] + p_1 [v]  v^\nu \nabla_\mu \del_\nu |v| )\right)+ \op (\lambda^3)
\end{multline}
with
\bne \zeta^\mu [v] = 2 {P^\mu}_\nu [\hat v] v^\sigma ({K^\nu}_\rho [\hat v] \nabla_\sigma \hat v^\rho - {R^\nu}_\sigma ) \ene
and
\bne p_1 [v] = \frac{2 k^{\mu\nu} [\hat v] (R_{\mu\nu} - {k_\mu}^\rho [\hat v] k_{\rho\nu} [\hat v] )}{v^\mu v^\nu \nabla_\mu \del_\nu |v|} .\ene

The bit thread formulation of GB HEE simplifies when the RT surface $m_0^*$ has no extrinsic curvature, such that $(K^{ij})^{(0)} = 0$ and so $(p_1 [v])^{(0)}$ vanishes on $m$, and we can choose $p_1 = 0$ to simplify the norm bound. The tangential component of $v_1$, $(\zeta^i [v])^{(0)}$ also vanishes on $m$ using the identity that relates the $iz$ component of the Ricci tensor to the vanishing extrinsic curvature,
\bne R_{iz} = \nabla_j {K^j}_i - \del_i K.\ene
Furthermore, a simpler form for $r$ can be used,
\bne r [\hat v] = R - 2 R_{\mu\nu} \hat v^\mu \hat v^\nu \ene
which still satisfies $(r [\hat v])^{(0)} = \tilde R^{(0)}$, $(r [\hat v])^{(1)} = \tilde R^{(1)}$ on $m$, so is adequate in giving the correct flux up to second order in $\lambda$. This gives us the norm bound
\bne \label{eq:bottleneck norm bound no extrinsic curvature} |v| \leq 1 + \lambda (R - 2R_{\mu\nu}\hat{v}^\mu \hat{v}^\nu ) + \op (\lambda^3) \ene
for cases where $m_0^*$ has no extrinsic curvature. We will compare this norm bound with the result derived using Lagrange dualization and find agreement.
\subsubsection{Gibbons-Hawking-York term}\label{sec:GHY}

The GHY term in $a_\lambda (m)$ has so far been neglected. We present two ways to incorporate it: adding a term to the norm bound with delta-function support on the boundary, and a doubling trick, taking $\del M$ to be the boundary both of the original Riemannian manifold $M$ and an identical copy, with bit threads flowing out into both.

The GHY term contains $\tilde K[\tilde n]$, the divergence of the surface's boundary normal $\tilde{n}$. By allowing an additional flux through on $\del m$ we capture the GHY term, however the difficulty is doing so without making reference to any surface. While we do not a priori know where the bottleneck will be, we do know what $v$ will be on it from which we can extract $\tilde n$ and thus $\tilde K [\tilde n ]$.

\begin{figure}[h]
\centering
\includegraphics[width=.5\textwidth]{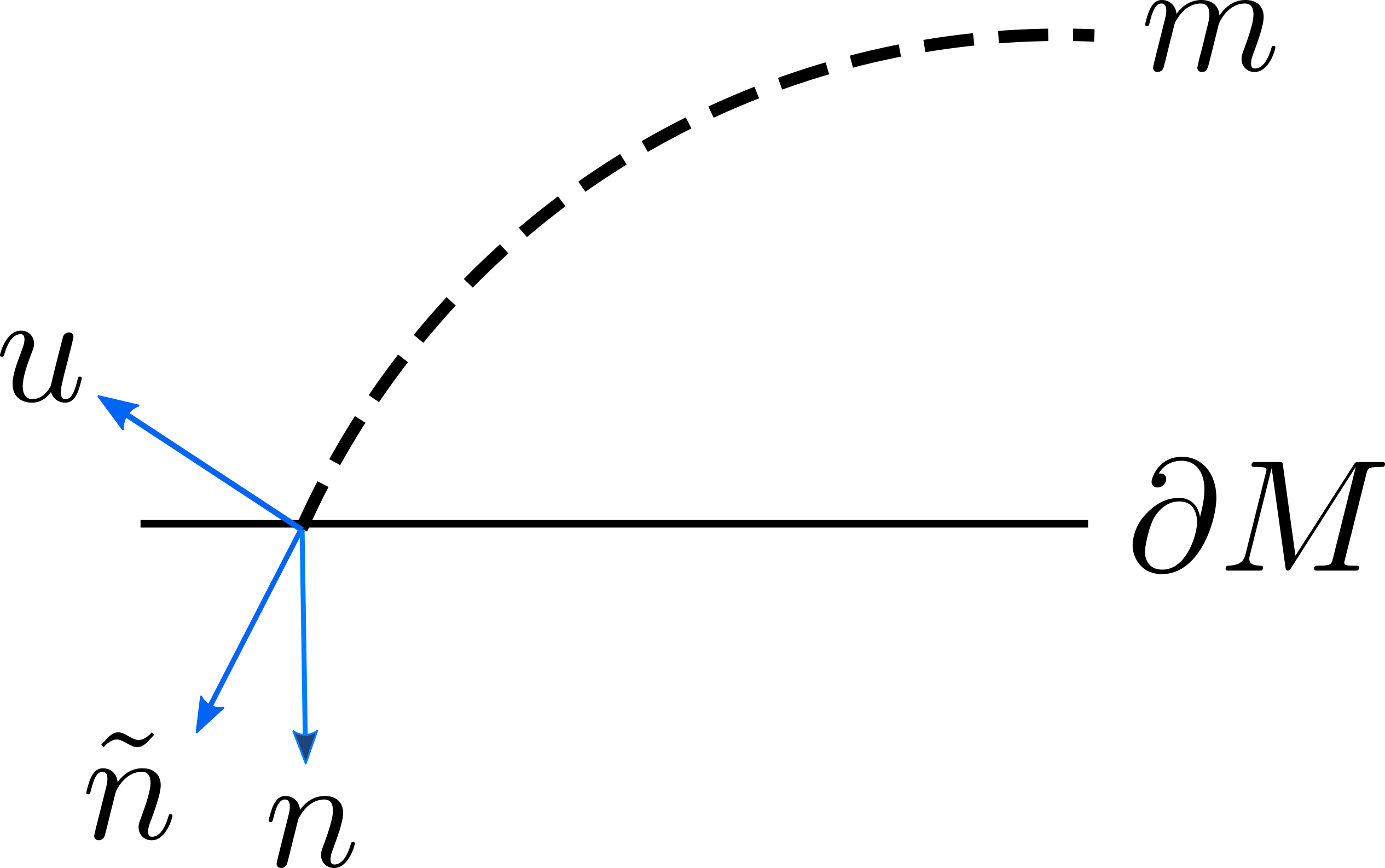}
\caption{Illustration showing $u$ the unit normal to $m$, $\tilde{n}$ the unit normal to $\del m$, and $n$ the unit normal to the bulk time slice boundary $\del M$ at $\del m$.}
\label{fig:vector diagram}
\end{figure}

Straightforwardly the unit normal $\tilde n$ can be written as the normalized projection of the time slice's boundary unit normal $n$ onto the tangent space of $m$, which is 
\bne \tilde{n}^\mu = \frac{{P^\mu}_\nu [u] n^\nu}{|P[u]n|}, \ene
see figure~\ref{fig:vector diagram}. We would like to adapt this formula for $\tilde n$ to use $v$ instead of $u$, therefore not making reference to any particular surface. Note also that on the boundary of $m$, as argued earlier the extrinsic curvature always vanishes, hence so too does $v_1^i$,
\bne \begin{split} v_1^i &= (\zeta^i [v])^{(0)}\\
 &= 2({k_i}^j [\hat v] \nabla_z \hat v _j - R_{iz}) ^{(0)} \\
&= 2({K_i}^j \nabla_z \hat v _j - \nabla_j {K^j}_i) ^{(0)} \\
&= 0 
 \end{split}\ene
and therefore $v$ is normal to $m$ to at least second order,
\bne \label{eq:v is very nearly perpendicular} v|_{\del m} = u + \op (\lambda^2) .\ene
Let us define a function on the spacetime boundary $\del M$
\bne \tilde{N}^\mu [v] \coloneqq \frac{{P^\mu}_\nu [v] n^\nu}{|P[v]n|} \ene
then at $\del A$ \eqref{eq:v is very nearly perpendicular} holds and so
\bne
\tilde{N}[v] = \tilde{n} + \op (\lambda^2)
\ene
on $\del A$.
To account for the GHY term we add to the norm bound a function with $\delta-$function support on $\del A$, which allows additional finite flux through $\del A$ equal to the GHY term
\bne \label{normal bound with GHY}
|v| \leq 1 + \lambda (f_1 [v]
+ 2 {k^\nu}_\nu[\tilde{N}[v]] \delta_{\del A}) + \lambda^2 f_2[v] + \op (\lambda^3)
\ene 
Note that the norm bound \eqref{normal bound with GHY} is defined purely in terms of boundary geometric data and an unconstrained $v$, such that bit thread thickness is only a function of local geometry and thread orientation. Using the formula for $\tilde N [v]$, our method generalises to any higher curvature HEE prescriptions whose boundary term is a functional of the surface boundary normal $\tilde n$. 

An alternative way of including the contribution of the GHY term is to employ a doubling trick. Taking $m$ to be a surface homologous to boundary subregion $A$, and adding the mirror image $\tilde{m}$ of $m$ across the boundary, creates a boundaryless surface $m+\tilde{m}$ for which
\bne \int_m \tilde{R} + 2 \int_{\del m} K = \frac{1}{2} \int_{m \cup \tilde{m}} \tilde{R}. \ene
After gluing the surface $m$ together with its double, there may be a kink in the surface at $\del A$, giving a singular induced scalar curvature. The GHY terms can then be understood as accounting for possible delta-function singularities in $\tilde{R}$ where we join $m$ with its mirror image. From the norm bound \eqref{normal bound with GHY}, this implies infinite bit thread density at $\del A$, though the flux is still finite. 

\begin{figure}[h]
\centering
\includegraphics[width=.5\textwidth]{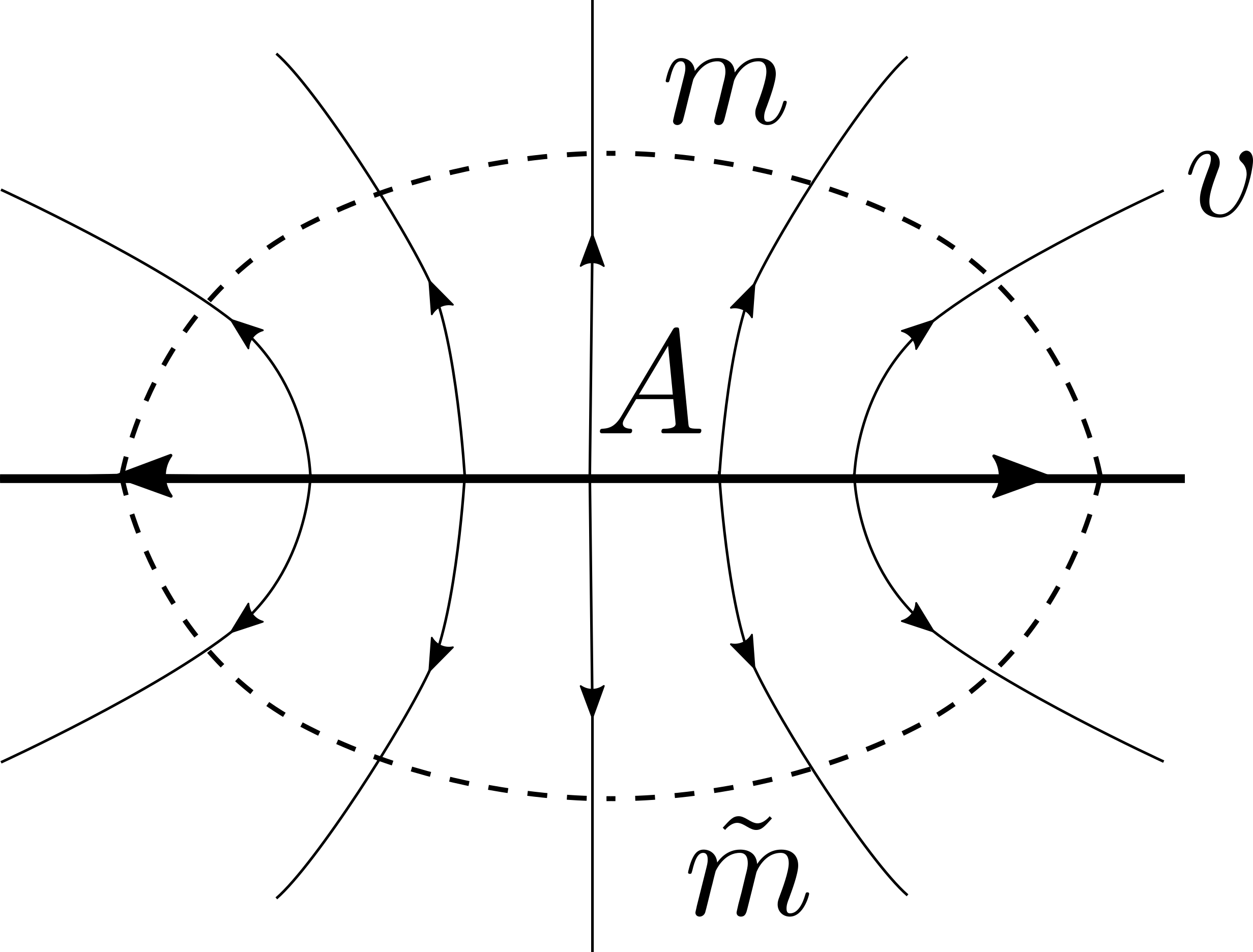}
\caption{By gluing two copies of the time slice $\Sigma$ along the boundary $\del \Sigma$, we create a boundaryless surface $m \cup \tilde{m}$. The GHY term is accounted for by the integral of $\tilde{R}$ over where $m$ and $\tilde{m}$ join.}
\label{fig:double_flow}
\end{figure}

In this doubling trick picture, the entanglement entropy is given by half the maximum flux out of boundary region $A$, 
\bne
4G_N S(A) = \frac{1}{2}\max_v \int_A v
\ene
where $v$ can flow out into two copies of $\Sigma$ glued along $\del \Sigma$, subject to divergenceless of $v$ and the norm bound, see figure~\ref{fig:double_flow}.

\section{Lagrange dualization method}\label{sec:Lagrange dualization}

The max flow-min cut theorem \eqref{eq:mfmc} is proven as a consequence of strong Lagrange duality between two convex optimization problems, namely max flow and a relaxed form of the min cut problem. (A review of these concepts aimed at physicists can be found in \cite{Headrick:2017ucz}.) In this section, we will apply these ideas to the Gauss-Bonnet holographic entanglement entropy formula. Unfortunately, as we will see, the $\lambda$ term in the funcational $a_\lambda(m)$ in general ruins the convexity of the relaxed min cut functional. Therefore, the technique will only work in certain special cases, namely when the minimal surface has no boundary and vanishing extrinsic curvature $K_{ij} = 0$, such as when calculating the entanglement entropy of one side in the high-temperature thermofield-double state. This will allow us to replace the non-convex optimization problem $\min\limits_{m\sim A} a_\lambda(m)$ with an equivalent convex optimization problem. This is important because non-convex problems generally have a duality gap between the primal and dual problem.\footnote{A general procedure exists called convex relaxation which allows one to embed a nonconvex problem in a larger solution space which is convex. When such a relaxation can be done it is possible to find a dual with zero duality gap. So far we have not been able to find such a relaxation which would allow the Gauss-Bonnet holographic entanglement entropy to be calculated in the general case. We leave this for future work.} Specializing to the situation where the problem is convex will then allow us to use Lagrange dualization to derive the flow reformulation.

\subsection{Convex optimization and Lagrange dualization}
We present here a brief review of the mathematics of Lagrange dualization and its application to HEE, however the authors strongly suggest that readers unfamiliar with these to read the more detailed expositions in sections 2 and 3 of \cite{Headrick:2017ucz} before trying to follow their extension to the higher-curvature case in subsection \ref{sec: HEE lagrange dualization} of this paper.

\subsubsection{Review of Lagrange dualization}
\label{sec:dualizationreview}

Lagrange duality is a technique often employed in the fields of linear programming and network theory. For a well defined class of minimization problems (the primal) there exists a description where the problem has been transformed into a maximization problem (the dual). Strong duality is the nontrivial assertion that these two descriptions are in fact the same, that the maximum of one equals the minimum of the other. 

Let $L_{p}$ and $\{f_a \}$ be a set of convex functions, and $\{h_b \}$ a set of affine functions on a vector space parametrized by $x$. The primal program is given by the constrained optimization program
\begin{equation}
\min_{x} L_{p}(x) \quad s.t.\quad f_{a}(x) \leq 0 \quad h_{b}(x)=0\,.
\end{equation}
We may rewrite $L_{p}$ by imposing Lagrange multipliers for the constraints
\begin{equation}
L(x, \{\phi_a \}, \{\gamma_b \}) \equiv L_{p}(x) + \phi_{a}f_{a}(x) + \gamma_{b}h_{b}(x), \quad \phi_{a} \geq 0.
\end{equation}
The primal problem $L_p$ may be recovered from $L$ by maximizing with respect to the Lagrange multipliers
\begin{equation}
L_{p}(x) = \max_{\phi_a \geq 0,\gamma_b} L(x, \{\phi_a \}, \{\gamma_b \}).
\end{equation}
To obtain the dual program however, we instead minimize $L$ with respect to $x$
\begin{equation}
L_{d}(\{\phi_a \}, \{\gamma_b \}) \equiv \min_{x} L\left(x, \{\phi_a \}, \{\gamma_b \}\right). 
\end{equation}
The difference between the solution of the primal program and the dual program is called the duality gap
\begin{equation}
d_g \equiv \min_{x}L_{p}(x_i) - \max_{\phi_a \geq 0,\gamma_b }L_{d}(\{\phi_a \}, \{\gamma_b \}).
\end{equation}
When the duality gap is zero then strong duality is said to hold. A sufficient, but not necessary condition for strong duality to hold is for $L_p(x_i)$ to be a convex function, and there to exist an $x_i$ in the relative interior of its domain for which the constraints are satisfied; this is Slater's condition.

When the dual program has a unique optimal configuration $(\phi_a^*,\gamma_b^*)$, then the values of the Lagrange multipliers tell us how sensitive the optimal value is to small changes in the constraints. In other words, if we replace the constraint $f_a(x)\le0$ by $f_a(x)+\lambda\le0$, then to first order in $\lambda$ the optimal value changes by $\lambda\phi_a^*$. This is derived for example in \cite{Headrick:2017ucz}. In fact, by a slight generalization of that argument, the result holds even when we perturb the constraint by a function $\delta f_a(x)$: if the dual optimal configuration is unique and if $\delta f_a(x^*)$ has the same value for all primal optimal points $x^*$, then replacing $f_a(x)\le0$ by $f_a(x)+\lambda\delta f_a(x)\le0$ changes the optimal value by $\lambda\phi_a^*\delta f_a(x^*)+\mathcal{O}(\lambda^2)$. The same result also holds for concave programs: replacing $f_a(x)\ge0$ by $f_a(x)+\lambda\delta f_a(x)\ge0$ changes the optimal value by $\lambda\phi_a^*\delta f_a(x^*)+\mathcal{O}(\lambda^2)$.
We will use this fact in subsection \ref{sec:perturbing}.

\subsubsection{Example: max flow-min cut} \label{sec:MFMC}

As an example to further familiarize readers, and to set the stage for the derivation with GB gravity, we will show how to apply Lagrange dualization to the RT formula with EH gravity, proving the Riemannian MFMC theorem. Let $M$ be a Riemannian manifold with boundary, in this case a constant time slice of a bulk spacetime in a static state of a holographic theory. Given a region $A \subset \partial M$ the HEE is
\begin{equation} 
S(A) = \frac1{4G_{\rm N}}\min_{m\sim A}\int_m \sqrt{\tilde g}\,.
\end{equation}
To define the problem as a well posed convex program we perform a convex relaxation of the program by adding a scalar field degree of freedom $\psi$, which is subject to the boundary condition $\psi |_{\del M} = \chi_A$ with $\chi_A = 1$ on $A$ and $0$ on the complement $A^c$. This has the effect of smearing the surface to form level sets of constant $\psi$ in the bulk. The optimal solution involves stacking these level sets all on the true minimal surface. The space of surfaces $m$ is a subspace of possible $\psi (x)$, when $\psi(x)$ is binary valued, equal to $1$ is a bulk region (not necessarily connected) and $0$ in the complement, then surface $m$ can be understood as the boundary of these regions. The resulting optimization problem is convex in $\psi$:
\begin{equation}\label{eq:relaxedmincut}
\min_{m\sim A} \int_m \sqrt{\tilde g} = \min_\psi \int_{M} \sqrt{g} \lvert \partial_{\mu}\psi \rvert, \quad \psi|_{\del M} = \chi_A\,.
\end{equation}
In order to proceed with the dualization we introduce Lagrange multiplier term $v^{\mu}$ enforcing the replacement of $\del_\mu \psi$ with a new vector degree of freedom $w^\mu$, and a boundary term that is minimized when $\psi|_{\del M} = \chi_A$
\begin{equation}
S(A) = \frac{1}{4G_N} \min_{\psi, v^\mu} \left [\int_{M}\sqrt{g}\,\left[\lvert w \rvert + v^{\mu}(w_{\mu}-\partial_{\mu}\psi)\right]+\int_{\partial M}\sqrt{h} \, |\psi-\chi_{A}| \right ].
\end{equation}
Optimizing first with respect to the Lagrange multipliers imposes the constraints and returns us to the primal program so let us instead optimize over the fields $w_{\mu}$ and $\psi$, giving the dual program
\begin{equation}\label{eq:EH gravity flow prescription}
S(A) = \frac{1}{4G_N} \max_{v^\mu} \int_{A}\sqrt{h}\,n_{\mu}v^{\mu},\quad \lvert v \rvert \leq 1 , \quad \nabla_{\mu}v^{\mu}=0.
\end{equation}
As the primal problem was convex and obeys Slater's condition, strong duality holds, and thus the Riemannian MFMC theorem is proven.

\subsubsection{Perturbing the convex program}
\label{sec:perturbing}

The flow formulation of RT HEE given by \eqref{eq:EH gravity flow prescription} is a well-posed convex program (more precisely, concave program, since it involves maximizing a concave functional). Dualizing it returns us to the relaxed min cut program \eqref{eq:relaxedmincut}. (See \cite{Headrick:2017ucz} for the details of this derivation.) By viewing the max flow program as the primal, we can use the relation between perturbations of the primal constraints and changes in the optimal value, described at the end of subsection \ref{sec:dualizationreview} above, to figure out how to change the norm bound in the max flow program in order to reproduce the $\lambda$ term in the GB HEE functional. This gives a very straightforward way to find the first-order correction to the norm bound.

In the dualization of the max flow program, there is a Lagrange multiplier $\psi$ for the divergencelessness constraint and another one $\phi$ for the norm bound. As long as the minimal surface $m_0^*$ is unique, the dual optimal configuration is also unique; in particular, $\phi^*$ is a delta function on $m_0^*$.

In this subsection we will only work to first order in $\lambda$. If we perturb the norm bound to make it
\begin{equation}
1-|v|+\lambda F\ge0\,,
\end{equation}
where $F$ is some function on $M$, then the maximum flux will change by
\begin{equation}
\lambda\int\sqrt{g}\,F\phi^* = \lambda\int_{m_0^*}\sqrt{\tilde g}F\,.
\end{equation}
In particular, if we choose $F$ to be any function which equals $\TCA$ on $m_0^*$, then the maximum flux will equal
\begin{equation}
\int_{m_0^*}\sqrt{\tilde g}\,(1+\lambda\TCA) = a_\lambda(m_0^*) = a_\lambda(m_\lambda^*)+\mathcal{O}(\lambda^2)\,,
\end{equation}
where we used the fact that $m_0^*$ extremizes the area, so $\area(m_\lambda^*)=\area(m_0^*)+\mathcal{O}(\lambda^2)$. In order for $F$ to equal $\TCA$ on $m_0^*$ for any max flow, we set it equal to $\tilde a [\hat v]$. The norm bound is thus
\begin{equation}
|v|\le1+\lambda \tilde a[\hat v]\,.
\end{equation}

\subsection{Lagrange dualization of higher curvature holographic entanglement entropy}
\label{sec: HEE lagrange dualization}

In this section, Lagrange dualization is applied to optimization problems of the form
\bne \min_{m \sim A}  \int_m \sqrt{\tilde{g}} (1 + \lambda \tilde{\mathcal{A}}), \ene
corresponding to a perturbative correction of the RT HEE prescription. 

We now carry out the same convex relaxation as in section~\ref{sec:MFMC}, such that the normal vector field with $\delta$-function support becomes a one-form $\del_\mu \psi$ supported over the bulk time slice $M$ with $\psi \in \mathbbm{R}$,
\bne u^\mu \to \frac{\del^\mu \psi}{|\del \psi|}. \ene Heuristically, this convex relaxation smears the surface over the manifold forming a foliation of hypersurfaces with $\del_\mu \psi / |\del \psi |$ the unit normal on a component surface. This gives
\bne \label{relaxed_Gauss_bonnet_EE}
\min_{\psi} \left [ \int_{M} \sqrt{g} \left(1+\lambda \tilde a \left[\frac{\del^\mu \psi}{|\del\psi|}\right]\right) |\del \psi | + \int_{\del M} \sqrt{h} \left(1+\lambda \tilde{\mathcal{A}}\right)| \chi_A -\psi | \right ] \ene
where $\chi_A =1$ in $A$ and $0$ in $A^c$. We restrict ourselves to the case where $\tilde{\mathcal{A}}$ depends on the surface unit normal $u$, but not derivatives of $u$ as they generally cause the problem to be non-convex. For example, suppose $\tilde{\mathcal{A}}$ contains terms involving the trace of the extrinsic curvature. $M$ is foliated by hypersurfaces of constant $\psi$, so smooth changes to $\psi(x)$ can lead to discontinous changes in hypersurface foliation, with very different extrinsic curvatures. A consequence of this is that the convexity condition 
\bne p K[\psi_a] + (1-p) K[\psi_b] \geq K[p \psi_a + (1-p) \psi_b] , \quad 0 \leq p \leq 1 \ene
can be violated to an arbitrary degree, making it a non-convex optimization problem.

We next add a Lagrange multiplier term $v_\mu (w^\mu - \del^\mu \psi)$ to replace derivatives of $\psi$ with $w^\mu$, arriving at the following Lagrangian:
\bne \begin{split} \label{eq:Lagrange problem}
L[\psi , w, v]= \int_M \sqrt{g}\left[(1 + \lambda \tilde a [\hat w])|w| + v_\mu (w^\mu - \del^\mu \psi)\right] \\ + \int_{\del M}\sqrt{h} (1 + \lambda \tilde{\mathcal{A}}) |\psi - \chi_A| \end{split}
\ene
where $\hat{w}^\mu \equiv w^\mu / |w|$.
We now minimize over the variables $\psi$ and $w^\mu$ on $M$ and $\del M$. Integrating the $v_\mu \del^\mu \psi$ term in \eqref{eq:Lagrange problem} by parts strips all derivatives off $\psi$, allowing us to do a pointwise minimization. The terms involving $\psi$ are
\bne 
\int_M \sqrt{g}\, \psi \nabla_\mu v^\mu + \int_{\del M} \sqrt{h}\left( (1 + \lambda \tilde{R})|\psi - \chi_A| + \psi n_{\del M}^\mu v_\mu\right). 
\ene
The bulk integrand is unbounded unless
\bne 
\nabla_\mu v^\mu = 0 \,,
\ene
in which case it vanishes. On the boundary, in order to have a bounded minimum in $\psi$, we require
\bne 
|n^\mu v_\mu| \leq 1 + \lambda \tilde{\mathcal{A}}\,; 
\ene
this inequality allows for the possibility that $A$ itself is the flow bottleneck. The minimum is at $\psi = \chi_A$, leaving us with
\bne \begin{split} \label{whateveR}
\min_\psi L[\psi, w, v]= \int_M \sqrt{g} \,\left[(1 + \lambda \tilde a [\hat w]  )|w| + v_\mu w^\mu\right] + \int_{A}\sqrt{h} v_\mu n_{\del M}^\mu \end{split} \ene
Let us minimise the bulk integrand
\begin{equation}
\left(1 + v_\mu\hat w^\mu+\lambda \tilde a [\hat w] \right)|w|\,.
\end{equation}
with respect to $w$. If the prefactor for $|w|$ is negative for any value of its direction $\hat w$, then the minimum is unbounded by sending its magnitude $|w| \to \infty$. Thus we require, for all values of $\hat w$, 
\bne \label{eq:v_inequality1} 
1 + v_\mu \hat{w}^\mu + \lambda  \tilde{\mathcal{A}}(\hat w) \geq 0\,,
\ene
and then the minimum is zero at $|w| = 0$. To see whether \eqref{eq:v_inequality1} holds for any $\hat w$, we minimize the left-hand side of the inequality with respect to $\hat w$, subject of course to the constraint $\hat w^\mu \hat w_\mu = 1$, finding the minimizing value for $\hat w$
\bne \begin{split} 
\label{the normal vector is v normalized}
\hat{w}^\mu &= - \frac{v^\mu + \lambda \frac{\del \tilde a [\hat w] }{\del \hat w_\mu}}{|v + \lambda \frac{\del \tilde a [\hat w]}{\del \hat w}|} \\
&= -\hat v^\mu - \frac{\lambda}{|v|} {P^\mu}_\nu [\hat{v}]  \frac{\del \tilde a [\hat w]}{\del \hat w_\nu}+ \lambda^2 \frac{(\hat{v}^\mu {P^\rho}_\nu [\hat{v}] + 2 \hat{v}^\rho {P^\mu}_\nu [\hat{v}]) }{2|v|^2} \frac{\del \tilde a [\hat w]}{\del \hat w_\nu} \frac{\del \tilde a [\hat w]}{\del \hat w^\rho}+ \op(\lambda^3) \end{split} 
\ene
using \eqref{eq:v hat lambda expansion}. Contracting with $v_\mu$ gives
\bne \label{eq:v contracted with w} v_\mu \hat w^\mu = - |v| + \frac{\lambda^2}{2|v|} P_{\mu\nu}[\hat v] \frac{\del \tilde a[\hat w]}{\del \hat w_\mu} \frac{\del \tilde a[\hat w]}{\del \hat w_\nu}.  \ene
Taking this minimizing value of $\hat w$ and Taylor expanding the $\lambda \tilde a[\hat w]$ term in \eqref{eq:v_inequality1} about $-\hat v$ gives
\bne \label{eq:taylor expansion of A} \lambda \tilde a[\hat w] = \left. \lambda \tilde a[\hat w]\right |_{\hat w = - \hat v} - \frac{\lambda^2}{|v|} P_{\mu\nu} [\hat v] \left. \left ( \frac{\del \tilde a[\hat w]}{\del \hat w_\mu}   \frac{\del \tilde a[\hat w]}{\del \hat w_\nu} \right) \right |_{\hat w = - \hat v} \ene
Substituting \eqref{eq:v contracted with w} and \eqref{eq:taylor expansion of A} into the  inequality \eqref{eq:v_inequality1} gives a constraint on $v$, which is the norm bound
\bne \label{modulus of v constraint with m}
|v| \leq 1 + \left. \lambda \tilde a[\hat w] \right |_{\hat w = - \hat v}- \frac{\lambda^2}{2} P_{\mu\nu} [\hat{v}] \left. \left ( \frac{\del \tilde a[\hat w]}{\del \hat w_\mu}   \frac{\del \tilde a[\hat w]}{\del \hat w_\nu} \right) \right |_{\hat w = - \hat v}+ \mathcal{O} (\lambda^3).
\ene
Bringing the constraints we have found together, we arrive at the dual problem
\bne \begin{split} \label{eq:dual_problem1} \max_v \int_{A} \sqrt{h} v_{\mu}n^\mu \text{ over } \{v^\mu : |v| < 1 + \left. \lambda \tilde a[\hat w] \right |_{\hat w = - \hat v}- \frac{\lambda^2}{2} P_{\mu\nu} [\hat{v}] \left. \left ( \frac{\del \tilde a[\hat w]}{\del \hat w_\mu}   \frac{\del \tilde a[\hat w]}{\del \hat w_\nu} \right) \right |_{\hat w = - \hat v}+ \mathcal{O} (\lambda^3) , \\ \nabla^\mu v_\mu = 0 , |n^\mu v_\mu | \leq (1 + \lambda \tilde{\mathcal{A}}). \}.\end{split}. \ene
Starting with the dual problem~\eqref{eq:dual_problem1}, one can reverse the process and recover the HEE formula~\eqref{gauss_bonnet_EE_altered}. The details of this calculation are non-essential to the conclusions of this paper, but it is worthwhile to note that from a convex maximal flow problem, one can find a dual minimal cut problem. 

\subsubsection{Application to Gauss-Bonnet gravity}
\label{sec: GB HEE lagrange dualization}

Here we will apply the results of the previous section to GB HEE, in the special case where the minimal surface has vanishing extrinsic curvature and no boundary, for which the minimization of surfaces becomes a convex problem. There are some non-convex optimization problems whose Lagrange dual obtains strong duality, GB HEE is not one of them. 

The GHY term contains the trace of the extrinsic curvature and is not convex, so we consider only surfaces without boundaries, for which the GB HEE formula is
\bne \label{gauss_bonnet_EE}
S(A) = \frac{1}{4G_N} \int_{m_\lambda^*} \sqrt{\tilde g}\left(1+ \lambda (R - 2R_{\mu\nu} u^\mu u^\nu + K^2 - K_{\mu\nu} K^{\mu\nu})\right).
\ene
As before, $m_\lambda^*$ is the codimension-2 surface homologous to $A$ that minimizes the surface functional, and $\tilde{g}_{\mu\nu}$ and $\tilde{R}$ are the induced metric and curvature scalar on $m_\lambda^*$. The extrinsic curvature terms in \eqref{gauss_bonnet_EE} are problematic to obtaining strong duality as they make the problem non-convex.

We will restrict ourselves to $m_0^*$ having no extrinsic curvature, then the $\lambda K^{\mu\nu} K_{\mu\nu}$ term in the GB HEE functional will be third order on $m_\lambda^*$ and can be dropped as we are only working to second order. The extrinsic curvature tensor appears only quadratically in GB HEE, so under the assumption that $m_0^*$ has no extrinsic curvature these terms can be removed without affecting the local minimum of~\eqref{gauss_bonnet_EE}. In cases where $m_0^*$ has vanishing curvature due to Killing symmetries, such as on bifurcation surfaces of Killing horizons, then $m_\lambda^*$ may also have vanishing extrinsic curvature. This is the case for all known static black hole event horizons in Lovelock gravity \cite{Jacobson1993}. 

 Thus we can take
\bne \label{gauss_bonnet_EE_altered}
S(A) = \frac{1}{4G_N} \min_{m\sim A} \int_{m} \sqrt{\tilde g}\left(1+ \lambda (R-2R_{\mu\nu}u^\mu u^\nu)\right)
\ene
as the primal program to dualize. We identify
\bne \tilde{\mathcal{A}} = R - 2 R_{\mu\nu} u^\mu u^\nu \ene
as the perturbation to the RT area functional, for cases where the optimum surface $m_\lambda^*$ has no boundary or extrinsic curvature. Following the procedure given in the previous section, after convex relaxation and substitution of $\psi$ with $\hat w$, this becomes
\bne \tilde{\mathcal{A}} (\hat w ) = R - 2 R_{\mu\nu} \hat w^\mu \hat w^\nu .\ene
for which, applying the result \eqref{modulus of v constraint with m}, gives the norm bound 
\bne |v| \leq 1 + \lambda (R - 2 R_{\mu\nu} \hat v^\mu \hat v^\nu ) - 8 \lambda^2 P^{\mu\nu} [\hat v]R_{\mu\rho} R_{\nu\sigma}\hat{v}^\rho \hat v^\sigma + \op (\lambda^3 ). \ene 
In fact, the $\op(\lambda^2)$ term in the above norm bound can be removed as they vanish on $m_\lambda^*$,
which follows from the vanishing of extrinsic curvature terms in the identity
\bne
R_{iz} = \nabla_j K^j_i - \partial_i K\,.
\ene
Thus the norm bound is simply
\bne \label{eq:LD new norm bound} |v| \leq 1 + \lambda ( R- 2 R_{\mu\nu} \hat v^\mu \hat v^\nu ) + \op (\lambda^3 ). \ene

There is perfect agreement between the norm bound found using Lagrange dualization \eqref{eq:LD new norm bound} and the norm bound found using the bottleneck method \eqref{eq:bottleneck norm bound no extrinsic curvature} in their overlapping regimes of validity: when $m_0^*$ has no boundary or extrinsic curvature. The non-trivial part of the agreement is that the second-order correction to the norm bound derived using the two methods both vanish. 

\section{Maximization over bit thread paths}
\label{sec: threadpaths}

There are special cases in which the corrected norm bound takes the form $|v|\le F_\lambda[\hat v]$, with the right-hand side depending only on the direction of $v$. An example is the one discussed at the end of the previous section, in which the unperturbed minimal surface $m_0^*$ has no extrinsic curvature, and the norm bound is given by \eqref{eq:LD new norm bound}. 
This suggests a decoupling of the norm $|v|$ and direction $\hat v$ of the vector field. However, the two are coupled by the divergencelessness constraint $\nabla_\mu v^\mu=0$. Here we will show that one can nonetheless decouple the direction and norm. Thus the problem of maximizing the flow can be decomposed into two steps: for a given $\hat v$, maximum the norm $|v|$; then maximize over $\hat v$.

In the language of bit threads, the direction field $\hat v$ specifies the potential thread configurations, while the norm bound fixes the maximum density. 

Consider a particular thread originating from a boundary point $x^i \in A$. Define a path $x^\mu (x^i ,s)$ along the thread as the integral curve along $\hat{v}^\mu$: the solution to 
\bne \frac{d}{ds}x^\mu (x^i ,s) = \hat{v}^\mu, \ene
 with $x^\mu (x^i ,s=0)$ the boundary point. The claim is that given knowledge only of the direction field $\hat v$, and the fact that we want to maximize the flux through $A$, we can find the thread number density everywhere in the bulk, and hence know everything about $v$.

First we show that if we know the thread density at any point on the thread, we know it for the whole thread.
The divergencelessness of $v$ can be written as 
\bne \hat{v}^\mu \nabla_\mu \ln |v| = - \nabla_\mu \hat{v}^\mu \ene
Integrating this along the bit thread from the boundary at $s=0$ to a point $s = s'$ gives
\bne \label{eq:integrating mod v}
|v|_{(x^i,s')} = |v|_{(x^i,0)} \exp \left(- \int_0^{s'} \nabla_\mu \hat{v}^\mu ds \right)
\ene
From \eqref{eq:integrating mod v} we see that, in order for $|v|$ to be single-valued, any loops of bit threads must obey $\oint \nabla_\mu \hat{v}^\mu = 0$. In fact, as any loops of bit threads in the bulk can only impede threads leaving $A$ and contribute nothing to the flux, we can assume without loss of generality that the direction field is free of loops. (Given a direction field containing loops, we can simply set $v$ to 0, making $\hat v$ undefined, on every point through which a loop passes.)

Next, we use the fact that, in order to maximize the flux out of $A$, for each point $x_i \in A$ we should increase $|v|_{(x_i,0)}$ until there is a point along the bit thread which saturates the norm bound, which occurs for
\bne \label{eq:width of individual bit thread}
|v|_{(x^i,0)} = \min_{s'} F_\lambda[\hat{v}]_{(x^i,s)} \exp \left(\int_{0}^{s'} \nabla_\mu \hat{v}^\mu ds \right).
\ene
Thus $|v|_{(x^i,0)}$ is known, which in turn tells us the thread density everywhere. 

Threads are always maximally packed on the minimal surface, and generally spread out towards the boundaries.~\eqref{eq:integrating mod v} says that when $\nabla_\mu \hat{v}^\mu < 0$ the threads are coming closer together, and when $\nabla_\mu \hat{v}^\mu > 0 $ the threads are moving apart. For RT bit threads, the minimal surface has $|v| = 1$ and hence $\nabla_\mu \hat{v}^\mu =0$ on it. In most of the bulk, the threads are free to come together or move apart, but in the neighborhood of either side of the minimal surface, there must be non-zero regions of $\nabla_\mu \hat{v}^\mu$, one side which is a source for the direction field, and the other a sink. The minimal surface thus emerges in this direction field picture as the surface which separates the two source and sink regions. For GB bit threads, there is a correction to this: the minimal surface will not perfectly demarcate bands of source and sink regions, as $\nabla_\mu \hat{v}^\mu$ does not necessariy vanish on $m_\lambda^*$.

Suppose one has specified a direction field $\hat{v}$ and this gives a set of integral curves. Each integral curve has its own bottleneck, at the value of $s'$ for which the exponential factor in \eqref{eq:width of individual bit thread} is smallest. We increase the value of $|v|$ on the boundary until the norm bound is saturated at that $s'$. For general direction fields, the union of neighbouring integral curve's bottleneck points won't be continuous, more like a random set of points, but for the special direction fields which give $m_\lambda^*$ that union of points is in fact the continuous minimal surface we are looking for\fn{In that union there will generally also be other scattered points.}. This is another way of seeing how $m_\lambda^*$ appears in the bit thread picture. 
Finally, we note that while the higher curvature corrections to the bit threads were incorporated by altering the norm bound, there are equivalent alternatives. The bit thread prescription is simple and has few components to it, there are only three aspects the corrections can affect: the divergence of $v$, the norm bound, or the objective functional. By a change of variables, redefining $ v^\mu \to F_\lambda[\hat{v}] v^\mu $ we regain the constant norm bound $|v|\le1$ at the cost of replacing the divergencelessness condition with $\nabla_\mu v^\mu = -v^\mu\partial_\mu F_\lambda[\hat v]$ and the objective functional with $\int_A F_\lambda[\hat{v}] v$. This field redefinition exchanges bit threads whose thickness varies with position and orientation, but must end on the boundary with threads that have constant thickness, but can start and end in the bulk. We should emphasize that this is only a change of variables. Even though the divergencesslessness condition has changed, it has nothing to do with quantum corrections. The specific form of the divergence here forces the new threads to follow the same integral curve on which they are created, effectively adding thickness to the thread. A general quantum correction would also give rise to a corrected divergencelessness condition, but would presumably allow threads to be created in the bulk which would flow more independently of the threads around it.

\section*{Acknowledgements}
The work of M.H. was supported in part by the National Science Foundation through Career Award No.\ PHY-1053842 and in part by the Simons Foundation through \emph{It from Qubit: Simons Collaboration on Quantum Fields, Gravity, and Information} and through a Simons Fellowship in Theoretical Physics. A.R. was supported by DoE grant DE-SC0009987 and the Simons Foundation. J.H. was supported by the National Science Foundation under the IGERT: Geometry and Dynamics Award No.\ 1068620. A.R. would like to Alastair Grant-Stuart for useful discussions. M.H. and J.H. would like to thank the International Centre for Theoretical Sciences (ICTS) for hospitality in the early stages of this work during a visit for the \emph{US-India Advanced Studies Institute: Classical and Quantum Information}. We would also like to thank the MIT Center for Theoretical Physics for hospitality. 

\appendix
\section{First order of quadratic obstruction equation} \label{sec:Linear order quadratic obstruction equation} 
The first order in \eqref{eq:quadratic obstruction} is 
\bne
\begin{split} \label{eq:quadratic first order}
&(v_{\mu}\nabla^{2}_{z}v^{\mu}+\left(\nabla_{z}v^{\mu}\right)^{2} -F_{\lambda}\nabla^{2}_{z}F_{\lambda} -\left(\nabla_{z}F_{\lambda}\right)^{2})^{(1)} \\ 
&= \left(g_{\mu\nu}v^{\mu}\nabla_{z}^{2}v^{\nu}+g_{\mu\nu}\nabla_{z}v^{\mu}\nabla_{z}v^{\nu}\right)^{(1)}-(\nabla_{z}^{2}F)^{(1)} \\
&= g_{\mu\nu}^{(0)}\left(v^{\mu}\nabla^{2}_{z}v^{\nu}+\nabla_{z}v^{\mu}\nabla_{z}v^{\nu}\right)^{(1)}+g^{(1)}_{\mu\nu}\left(v^{\mu}\nabla^{2}_{z}v^{\nu}+\nabla_{z}v^{\mu}\nabla_{z}v^{\nu}\right)^{(0)}-(\nabla_{z}^{2}f_1[v])^{(0)}\\
&=g_{\mu\nu}^{(0)}\left(v_{0}^{\mu}(\nabla^{2}_{z}v^{\nu})^{(1)}+v_{1}^{\mu}(\nabla^{2}_{z}v^{\nu})^{(0)}+2(\nabla_{z}v^{\mu})^{(0)}(\nabla_{z}v^{\nu})^{(1)}\right)+\tilde{g}^{(1)}_{ij}(\nabla_{z}v^{i})^{(0)}(\nabla_{z}v^{j})^{(0)}-(\nabla_{z}^{2}f_1[v])^{(0)} \\
&=(\nabla^{2}_{z}v^{z})^{(1)}+v_{1}^{z}(\nabla^{2}_{z}v^{z})^{(0)}+ \tilde{g}_{ij}^{(0)} v_1^i (\nabla_z^2 v^j)^{(0)} + 2\tilde{g}_{ij}^{(0)}(\nabla_{z}v^{i})^{(0)}(\nabla_{z}v^{j})^{(1)}\\
&\qquad\qquad\qquad\qquad\qquad\qquad\qquad\qquad\qquad+\tilde{g}^{(1)}_{ij}(\nabla_{z}v^{i})^{(0)}(\nabla_{z}v^{j})^{(0)}-(\nabla_{z}^{2}f_1[v])^{(0)} \leq 0
\end{split}
\ene
When we reach the second-order calculation we will be performing a pointwise maximization with respect to $v_1^i$. If the above simplifies to a constraint purely on $v_1^i$ and its derivative tangential to $m$ then it reduces the space of feasible $v_1^i$ and will be important in the second-order calculation. However, if unconstrained variables such as $\del_z v_1^i$ do not vanish then this bound places no real constraint on the value of $v_1$ on $m$. We need to be especially careful to make full use of the divergenceless of $v$, which relates derivative of $v$ in different directions. The evaluation of each term in \eqref{eq:quadratic first order} gives
\bne
\begin{split}
\left(\nabla_{z}v^{i}\right)^{(0)} &= \partial_{z}v_{0}^{i}+{K^{(0)i}}_{j}v_{0}^{j} = \partial_{z}v_{0}^{i}\\
\left(\nabla_{z}v^{i}\right)^{(1)} &= \partial_{z}v_{1}^{i}+(\Gamma^i_{\mu z}v^\mu)^{(1)} = \partial_{z}v_{1}^{i}+{K^{(0)i}}_{j}v_{1}^{j} \\
\left(\nabla_{z}^{2}v^{i}\right)^{(0)}& = \partial_{z}(\partial_{z}v_{0}^{i}+{K^{(0)i}}_{j}v^{j}_{0})+{K^{(0)i}}_{l}(\partial_{z}v_{0}^{l}+{K^{(0)l}}_{j}v_{0}^{j}) = \del_z^2 v_0^i + 2{K^{(0)i}}_{j}\del_z v_0^j\\
\nabla_z^2 v^z &= \nabla_z (\nabla_\mu v^\mu - \nabla_i v^i ) = -\nabla_z \nabla_i v^i = - \del_i \del_z v^i - v^z \del_z K - v^j \del_z \Gamma^{i}_{ij} - K \del_z v^z - \Gamma_{ij}^i \del_z v^j \\
\left (\nabla_z^2 v^z \right )^{(0)} &= - \del_i \del_z v_0^i - \del_z K^{(0)} - \Gamma^{(0)i}_{ij}\del_z v_0^j \\
(\nabla_z^2 v^z )^{(1)} &= - \del_i \del_z v_1^i - \tilde{R}^{(0)} \del_z K^{(0)} - \del_z K^{(1)} - v_1^j \del_z \Gamma^{(0)i}_{ij} - \Gamma^{(0)i}_{ij} \del_z v_1^j - \Gamma^{(1)i}_{ij} \del_z v_0^j\\
(\nabla_{z}^{2}f_1 [\hat{v}])^{(0)} &= \del_z^2 f_1 [v_0]  
\end{split}
\ene
which allows us to write \eqref{eq:quadratic first order} as
\bne \label{eq:quadratic first order simplified}-\del_i \del_z v_1^i + A_i [v_0] \del_z v_1^i + B_i[v_0] v_1^i+ C[v_0] \leq 0, 
\ene
with the definitions
\bne \begin{split} \label{eq:definitions for A B and C}
A_i [v_0] &\coloneqq (2\tilde{g}_{ij}^{(0)} \del_z v_0^j - \Gamma^{(0)j}_{ij})\\
B_i [v_0] &\coloneqq 2 \tilde{g}_{ij}^{(0)}(\del_z^2 v_0^j + 4{K^{(0)j}}_l \del_z v_0^l- \del_z \Gamma^{(0)i}_{ij})\\
C[v_0] &\coloneqq - \del_z K^{(1)} - \Gamma^{(1)i}_{ij}\del_z v_0^j + \tilde{R}^{(0)}(-\del_i \del_z v_0^i - 2\del_z K^{(0)} - \Gamma^{(0)i}_{ij}\del_z v_0^j) +\tilde{g}_{ij}^{(1)} \del_z v_0^i \del_z v_0^j - \del^2 f_{1}[v_0].
\end{split} \ene 
Eqn \eqref{eq:quadratic first order simplified} contains $\del_z v_1^i$, so there is no real constraint on $v_1^i$ from this obstruction equation. In contrast the first-order linear obstruction equation has no such $\del_z v^i_1$ terms and the set of obstructionless $v_1^i$ which the flow maximizes over is generally a subset of all $v_1^i$. 

\bibliographystyle{JHEP}
\bibliography{Holographic_bit_threads_in_higher_curvature_gravity}

\providecommand{\href}[2]{#2}\begingroup\raggedright\begin{thebibliography}{10}

\bibitem{Ryu:2006bv}
S.~Ryu and T.~Takayanagi, \emph{{Holographic derivation of entanglement entropy
  from AdS/CFT}},
  \href{https://doi.org/10.1103/PhysRevLett.96.181602}{\emph{Phys. Rev. Lett.}
  {\bfseries 96} (2006) 181602}
  [\href{https://arxiv.org/abs/hep-th/0603001}{{\ttfamily hep-th/0603001}}].

\bibitem{Freedman:2016zud}
M.~Freedman and M.~Headrick, \emph{{Bit threads and holographic entanglement}},
  \href{https://doi.org/10.1007/s00220-016-2796-3}{\emph{Commun. Math. Phys.}
  {\bfseries 352} (2017) 407}
  [\href{https://arxiv.org/abs/1604.00354}{{\ttfamily 1604.00354}}].

\bibitem{Bhattacharyya:2013gra}
A.~Bhattacharyya, M.~Sharma and A.~Sinha, \emph{{On generalized gravitational
  entropy, squashed cones and holography}},
  \href{https://doi.org/10.1007/JHEP01(2014)021}{\emph{JHEP} {\bfseries 01}
  (2014) 021} [\href{https://arxiv.org/abs/1308.5748}{{\ttfamily 1308.5748}}].

\bibitem{Bhattacharyya:2013jma}
A.~Bhattacharyya, A.~Kaviraj and A.~Sinha, \emph{{Entanglement entropy in
  higher derivative holography}},
  \href{https://doi.org/10.1007/JHEP08(2013)012}{\emph{JHEP} {\bfseries 08}
  (2013) 012} [\href{https://arxiv.org/abs/1305.6694}{{\ttfamily 1305.6694}}].

\bibitem{Bhattacharyya:2014yga}
A.~Bhattacharyya and M.~Sharma, \emph{{On entanglement entropy functionals in
  higher derivative gravity theories}},
  \href{https://doi.org/10.1007/JHEP10(2014)130}{\emph{JHEP} {\bfseries 10}
  (2014) 130} [\href{https://arxiv.org/abs/1405.3511}{{\ttfamily 1405.3511}}].

\bibitem{Jacobson1993}
T.~Jacobson and R.~C. Myers, \emph{{Black hole entropy and higher curvature
  interactions}},
  \href{https://doi.org/10.1103/PhysRevLett.70.3684}{\emph{Phys. Rev. Lett.}
  {\bfseries 70} (1993) 3684}
  [\href{https://arxiv.org/abs/hep-th/9305016}{{\ttfamily hep-th/9305016}}].

\bibitem{Caceres2017}
E.~Caceres, M.~Sanchez and J.~Virrueta, \emph{{Holographic Entanglement Entropy
  in Time Dependent Gauss-Bonnet Gravity}},
  \href{https://doi.org/10.1007/JHEP09(2017)127}{\emph{JHEP} {\bfseries 09}
  (2017) 127} [\href{https://arxiv.org/abs/1512.05666}{{\ttfamily
  1512.05666}}].

\bibitem{Erdmenger2014}
J.~Erdmenger, M.~Flory and C.~Sleight, \emph{{Conditions on holographic
  entangling surfaces in higher curvature gravity}},
  \href{https://doi.org/10.1007/JHEP06(2014)104}{\emph{JHEP} {\bfseries 06}
  (2014) 104} [\href{https://arxiv.org/abs/1401.5075}{{\ttfamily 1401.5075}}].

\bibitem{Chen2013}
B.~Chen and J.-j. Zhang, \emph{{Note on generalized gravitational entropy in
  Lovelock gravity}},
  \href{https://doi.org/10.1007/JHEP07(2013)185}{\emph{JHEP} {\bfseries 07}
  (2013) 185} [\href{https://arxiv.org/abs/1305.6767}{{\ttfamily 1305.6767}}].

\bibitem{MohammadiMozaffar2016}
M.~R. Mohammadi~Mozaffar, A.~Mollabashi, M.~M. Sheikh-Jabbari and M.~H.
  Vahidinia, \emph{{Holographic Entanglement Entropy, Field Redefinition
  Invariance and Higher Derivative Gravity Theories}},
  \href{https://doi.org/10.1103/PhysRevD.94.046002}{\emph{Phys. Rev.}
  {\bfseries D94} (2016) 046002}
  [\href{https://arxiv.org/abs/1603.05713}{{\ttfamily 1603.05713}}].

\bibitem{Dong:2013qoa}
X.~Dong, \emph{{Holographic Entanglement Entropy for General Higher Derivative
  Gravity}}, \href{https://doi.org/10.1007/JHEP01(2014)044}{\emph{JHEP}
  {\bfseries 01} (2014) 044} [\href{https://arxiv.org/abs/1310.5713}{{\ttfamily
  1310.5713}}].

\bibitem{Camps2014}
J.~Camps, \emph{{Generalized entropy and higher derivative Gravity}},
  \href{https://doi.org/10.1007/JHEP03(2014)070}{\emph{JHEP} {\bfseries 03}
  (2014) 070} [\href{https://arxiv.org/abs/1310.6659}{{\ttfamily 1310.6659}}].

\bibitem{Haehl2017}
F.~M. Haehl, E.~Hijano, O.~Parrikar and C.~Rabideau, \emph{{Higher Curvature
  Gravity from Entanglement in Conformal Field Theories}},
  \href{https://arxiv.org/abs/1712.06620}{{\ttfamily 1712.06620}}.

\bibitem{deBoer:2011wk}
J.~de~Boer, M.~Kulaxizi and A.~Parnachev, \emph{{Holographic Entanglement
  Entropy in Lovelock Gravities}},
  \href{https://doi.org/10.1007/JHEP07(2011)109}{\emph{JHEP} {\bfseries 07}
  (2011) 109} [\href{https://arxiv.org/abs/1101.5781}{{\ttfamily 1101.5781}}].

\bibitem{Hung:2011xb}
L.-Y. Hung, R.~C. Myers and M.~Smolkin, \emph{{On Holographic Entanglement
  Entropy and Higher Curvature Gravity}},
  \href{https://doi.org/10.1007/JHEP04(2011)025}{\emph{JHEP} {\bfseries 04}
  (2011) 025} [\href{https://arxiv.org/abs/1101.5813}{{\ttfamily 1101.5813}}].

\bibitem{Headrick:2017ucz}
M.~Headrick and V.~E. Hubeny, \emph{{Riemannian and Lorentzian flow-cut
  theorems}}, \href{https://doi.org/10.1088/1361-6382/aab83c}{\emph{Class.
  Quant. Grav.} {\bfseries 35} (2018) 105012}
  [\href{https://arxiv.org/abs/1710.09516}{{\ttfamily 1710.09516}}].

\end{thebibliography}\endgroup

\end{document}